\documentclass[reprint, aps, amsmath, amssymb]{revtex4-2}

\usepackage{epsfig}
\usepackage{graphicx}   
\usepackage{dcolumn}    
\usepackage{bm}         
\usepackage{color}

\usepackage{lmodern} 
\usepackage[bookmarks, colorlinks=false]{hyperref}
\usepackage{mathtools}
\hypersetup{colorlinks=true, allcolors=blue}
\usepackage[english]{babel}

\def\dint#1{\text{d}#1}

\renewcommand{\vec}[1]{{\bf #1}}

\newcommand{\wmean}[0]{\overline{\omega}}

\usepackage{siunitx}

\usepackage[caption=false]{subfig}

\begin{document}

\title{Macroscopic elastic stress and strain produced by irradiation}
\author{Luca Reali}
\email{Luca.Reali@ukaea.uk}
\affiliation{UK Atomic Energy Authority, Culham Science Centre, Oxfordshire OX14 3DB, UK}

\author{Max Boleininger}
\email{Max.Boleininger@ukaea.uk}
\affiliation{UK Atomic Energy Authority, Culham Science Centre, Oxfordshire OX14 3DB, UK}

\author{Mark R. Gilbert}
\email{Mark.Gilbert@ukaea.uk}
\affiliation{UK Atomic Energy Authority, Culham Science Centre, Oxfordshire OX14 3DB, UK}

\author{Sergei L. Dudarev}
\email{Sergei.Dudarev@ukaea.uk}
\affiliation{UK Atomic Energy Authority, Culham Science Centre, Oxfordshire OX14 3DB, UK}

\begin{abstract}
Using the notion of eigenstrain produced by the defects formed in a material exposed to high energy neutron irradiation, we develop a method for computing macroscopic elastic stress and strain arising in components of a fusion power plant during operation. In a microstructurally isotropic material, the primary cause of macroscopic elastic stress and strain fields is the spatial variation of neutron exposure. We show that under traction-free boundary conditions, the volume-average elastic stress always vanishes, signifying the formation of a spatially heterogeneous stress state, combining compressive and tensile elastic deformations at different locations in the same component, and resulting solely from the spatial variation of radiation exposure. Several case studies pertinent to the design of a fusion power plant are analysed analytically and numerically, showing that a spatially varying distribution of defects produces significant elastic stresses in ion-irradiated thin films, pressurised cylindrical tubes and breeding blanket modules.

Keywords: Irradiation-induced stress, swelling, elasticity, neutron irradiation, ion irradiation, multi-scale modelling

\end{abstract}

\maketitle

\section{Introduction}

The accumulation of microscopic defects resulting from the exposure of materials to neutron irradiation is one of the critical factors limiting the lifetime of components in a fission power plant \cite{Odette2009} and, by implication, is an issue of fundamental significance in the context of a comparative assessment of designs of power-generating fusion reactors \cite{Stork2014,Federici2019}. The ambitious timescale for the development of fusion reactors forces one to rely on virtual engineering \cite{Thompson2011,Barrett2018}, hence creating a strong drive towards the development of the so-called ``digital twins'', a relatively new concept in the area of nuclear fusion \cite{Rauscher2021}. It is important to include in this process the consideration of operating conditions, projecting the reactor design beyond the digital twin concept, to enable the assessment of effects of neutron irradiation. This point was highlighted by Jumel {\it et al.} \cite{Jumel2000} in relation to the problem of ageing of fission reactors. The present study and our earlier work \cite{NF2018} are the initial steps in establishing the fundamental link between the microscopic effects of exposure of materials to irradiation and macroscopic component-scale finite element virtual reactor models.

In a fission reactor, cladding materials receive the highest exposure to neutron irradiation \cite{Holt1988,Adamson2019}, comparable to the radiation exposure expected in a fusion power plant. The accumulation of defects in a zirconium alloy, a material often used as cladding for uranium oxide fissile nuclear fuel, gives rise to anisotropic dimensional changes and irradiation-induced deformations in the limit of high irradiation dose. Since zirconium alloys have hexagonal close-packed (hcp) crystal structure, the observed dimensional changes stem primarily from the texture of the cladding material \cite{Preuss2010,Adamson2019} and the intrinsic anisotropy of the hcp crystal structure. Not only the point defects in hcp crystals are anisotropic \cite{Domain2005}, but the dynamics of agglomeration of defects into mesoscopic and macroscopic dislocation loops and dislocation networks is anisotropic as well \cite{Preuss2017,Adamson2019}, resulting in significant macroscopic deformations developing in the limit of high irradiation dose \cite{Holt1988,Adamson2019}.  

On the other hand, the candidate materials typically selected for the structural components of a fusion power plant all have isotropic cubic crystal structure. It is either the body-centred cubic (bcc) structure characterising iron-chromium alloys, ferritic steels and tungsten, or the face-centred cubic (fcc) structure of copper, copper alloys and austenitic steels \cite{Stork2014,Marian2017,Pintsuk2019,Henry2019,Rieth2021}. Exposure of these materials to irradiation gives rise to swelling or, in other words, to the predominantly isotropic volume expansion resulting from the accumulation of defects in the material \cite{Zinkle2011,Zinkle2014,Garner2020,Klimenkov2021}. In the absence of applied load or other constraints, no significant anisotropic dimensional changes are expected to occur in the crystallographically isotropic materials selected for the structural components of a fusion power plant. 

A remaining source of anisotropy stems from the macroscopic spatial variation of the neutron exposure itself. Components of a fusion power plant are going to be exposed to neutron and ion irradiation generated by a localised source, the fusion plasma. The flux of high energy 14.1~MeV neutrons attenuates in the blanket surrounding the plasma on the length scale of several tens of centimetres \cite{Sato2003,Gilbert2013}. Energetic ions penetrate into the plasma-facing materials to the depth of several micrometres \cite{Markelj2019,Mason2020,Mason2021}, whereas hydrogen and helium produced either by transmutation or by a direct exposure to the fusion plasma, can diffuse deep into the bulk of structural components, particularly at elevated temperatures \cite{Shimada2019}.   

Anisotropic stress in reactor components stems not only from the formation of radiation defects. Gravitational body forces act in the direction of ${\bf g}$ and the magnitude of gravitational stress is appreciable, for example the 54 divertor cassettes in ITER weighs approximately 10 tons each~\cite{raffray2015}. The magnetic field in a tokamak exerts magnetic body forces on the ferromagnetic ferritic-martensitic steels in the breeding blanket modules, and these forces are expected to be several times stronger than gravity. Thermal expansion is also expected to generate high stresses in reactor components \cite{Bonelli2015}. The difference between the forces from gravity, magnetic fields and thermal expansion, and the internal forces from the defects formed following the exposure of materials to neutron irradiation is that irradiation changes the microscopic structure of a material, and hence the effects of irradiation, including radiation-induced deformations and stress, are in most cases irreversible.  

The formation of radiation defects \cite{MaPRM2019b,MaPRM2021} and defects associated with gas impurity atoms \cite{Hofmann2015} give rise to the volumetric expansion of the material \cite{Mason2019}. The magnitude of this volumetric expansion varies depending on the irradiation dose and temperature \cite{Garner2020}, whereas the presence of helium and hydrogen may have a significant additional non-linear effect on the microstructure and swelling rates \cite{Maziasz1982,Sandoval2015,Sandoval2018}. 

Swelling typically increases monotonically with exposure to irradiation and exhibits a complex pattern of variation as a function of dose and temperature \cite{Allen2015,Garner2020}. It is natural to expect that in an operating reactor, where different parts of every component have different temperatures and are exposed to irradiation at different dose rates, swelling is spatially non-uniform. In this study, we investigate what effect this spatial non-uniformity of swelling is going to have on the stress and strain developing in reactor components during the operation of the reactor. The non-uniform irradiation, temperature and loading conditions developing in large-scale reactor components are different from the uniform or zero-stress state developing in test samples of materials employed in the context of irradiation experiments and tests \cite{Pintsuk2019,Rieth2021b}.

In an earlier study \cite{NF2018}, we found that the density of relaxation volumes of defects represents the fundamental notion linking microscopic and macroscopic scales in a holistic simulation of radiation-induced deformations. The treatment developed in \cite{NF2018} was focused solely on the evaluation of the {\it total} stress and strain fields, which include both the elastic and defects' own contributions to deformation. 

However, only the elastic stress enters the engineering structural integrity criteria. Here, we develop an approach that enables computing the purely {\it elastic} stresses and deformations in an irradiated component on the macroscale, with an additional emphasis on the effect of boundary conditions.

The central concept that enables separating the purely elastic part of deformation from the total deformation, and at the same time developing a genuinely multiscale approach to the problem, is the notion of Mura's eigenstrain \cite{Mura}, and the fact that it can be identified uniquely with the tensorial density of relaxation volumes of defects. This statement and a proof of its validity is given in the first section of the paper.

We then explore several case studies involving representative geometries of components or samples exposed to irradiation. They include a specimen of rectangular geometry often used in the context of ion irradiation experiments. We then explore the deformations developing in a pressurised cylindrical pipe exposed to irradiation from an external source. This case describes a typical test sample used for investigating the combined effect of stress and neutron irradiation on materials. The most accessible to mathematical analysis case is a spherical shell exposed to a centrally-symmetric source of irradiation. This case admits an exact analytical solution, which agrees fully with numerical finite element model (FEM) simulations. Finally, using a purely numerical FEM approach, we evaluate elastic stress and strain developing in a cellular rectangular component exposed to irradiation. Such a cellular structure resembles a section of a tritium breeding blanket of a fusion tokamak reactor, and the simulations illustrate a complex pattern of stress and strain that is expected to develop in a geometrically complex macroscopic reactor component exposed to a spatially varying neutron flux -- entirely different from the cases where irradiation is more uniform.

In this work, we adopt the same approximation as in Ref. \cite{NF2018}, where the density of relaxation volume tensors of defects is taken as diagonal with respect to its Cartesian components. This spatially-varying isotropic volumetric swelling approximation is valid if the defects adopt random configurations with respect to their internal directional degrees of freedom \cite{NF2018}. There are cases where the applied stress, or the stress generated by the defects themselves, polarize the relaxation volume tensor density of defects \cite{Boleininger2019,Mason2020,Lai2020}. This aspect of microstructural evolution and its effect on the elastic stress and strain fields, while being accessible to the methodology developed below, is a non-linear microstructural relaxation phenomenon that remains outside the scope of this study. 

When investigating the various case studies below, we take the local relaxation volume density of defects at a given location as proportional to the radiation exposure of the material at that location. In a real operating scenario, the density of relaxation volumes at a given location is given by a complex convolution of histories of radiation exposure, temperature and stress at that location, see for example Refs. \cite{Derlet2020,Mason2020}, which themselves depend in a self-consistent manner on the history of evolution of these quantities at other locations in a component and possibly even in the entire reactor structure. Still, irrespectively of the approximations adopted for the evaluation of the density of relaxation volumes of defects, the fundamental link between this entity and the elastic strain and stress fields generated by irradiation remains unique, and it is this link and its implications that we establish and explore in detail below.

\section{Defect eigenstrain theorem}

Since the spatial scale of a macroscopic component is many orders of magnitude larger than the microscopic size of defects present in the material, the field of displacements ${\bf u}({\bf x})$ generated by a defect situated at ${\bf x}'$ can be evaluated using the far-field approximation as
%
%
\begin{equation}\label{eq:udisppt}
    u_i({\bf x}) = -P_{jl} {\partial \over \partial x_l}G_{ij}({\bf x}-{\bf x}'),
\end{equation}
where $P_{jl}$ is the elastic dipole tensor of the defect \cite{Leibfried1978,PRB2018}, and $G_{ij}({\bf x}-{\bf x}')$ is the Green function of elasticity equations \cite{Mura,Sutton2020}. A general form of equation \eqref{eq:udisppt}, describing the field of displacements produced by an ensemble of defects distributed in the material, is
\begin{equation}
    u_{i}({\bf x}) = -\int \Pi _{jl}({\bf x}'){\partial \over \partial x_l}
                    G_{ij}({\bf x}-{\bf x}') \dint{^3 x'},
\end{equation}
where $\Pi_{jl}({\bf x})=\sum_{a} P^{(a)}_{jl} \delta ({\bf x}-{\bf R}_a)$ is the density of elastic dipole tensors of defects. This field is related to another field $\omega _{mn}({\bf x})$, describing the density of relaxation volume tensors of defects, {\it via}
\begin{equation}
    \Pi_{jl}({\bf x}) = C_{jlmn}\omega_{mn}({\bf x}),
\end{equation}
where $C_{jlmn}$ is the fourth-rank tensor of elastic constants.
The density of relaxation volume tensors $\omega _{mn}({\bf x})$ is defined as
\begin{equation}\label{volume_tensor_density}
    \omega_{mn}({\bf x}) = \sum_{a} \Omega^{(a)}_{mn} \delta({\bf x}-{\bf R}_a), 
\end{equation}
where $\Omega^{(a)}_{mn}$ is the relaxation volume tensor of a defect situated at ${\bf R}_a$. Relaxation volume tensors of defects can now be routinely computed using density functional theory \cite{MaCPC2019,MaPRM2019a,MaPRM2019b,MaPRM2021} or, with lesser accuracy, using molecular dynamics simulations, where interactions between the atoms are described by semi-empirical potentials \cite{Mason2019,MaPRM2020}. For a particular defect configuration, the tensorial density of relaxation volumes is given by equation (\ref{volume_tensor_density}), whereas in the general case it can be treated as an ensemble-averaged quantity \cite{NF2018} 
\begin{equation}\label{volume_tensor_density_ensemble}
    \omega_{mn}({\bf x}) = \big\langle\sum_{a} \Omega^{(a)}_{mn} \delta({\bf x}-{\bf R}_a) \big\rangle, 
\end{equation}
where $\langle ...\rangle$ refers to averaging over a statistical ensemble of representative realisations of microstructure. Function  (\ref{volume_tensor_density}) is singular at locations occupied by the defects, whereas (\ref{volume_tensor_density_ensemble}) is non-singular.

In terms of the density of relaxation volume tensors of defects (\ref{volume_tensor_density}) and (\ref{volume_tensor_density_ensemble}), the field of displacements can be expressed as
\begin{equation}\label{eq:displacement_eigenstrain}
    u_i({\bf x}) = -\int C_{jlmn} \omega_{mn}({\bf x}'){\partial \over \partial x_l}
                    G_{ij}({\bf x}-{\bf x}') \dint{^3x'}.
\end{equation}
Comparing this relation with equation (3.23) by Mura \cite{Mura}
\begin{equation}
    u_i({\bf x}) = -\int C_{jlmn} \epsilon^*_{mn}({\bf x}'){\partial \over \partial x_l}
                    G_{ij}({\bf x}-{\bf x}') \dint{^3x'},
\end{equation}
we establish and prove the {\it defect eigenstrain theorem}, namely the statement that Mura's spatially-varying field of eigenstrains $\epsilon^*_{mn}({\bf x})$ is identical to the density of relaxation volume tensors of defects, viz.
\begin{equation}\epsilon^*_{mn}({\bf x})\equiv \omega _{mn} ({\bf x}).\label{defect_eigenstrain_theorem}
\end{equation} 

The recognition of equivalence of eigenstrain $\epsilon ^*_{mn}({\bf x})$ to the density of relaxation volume tensors $\omega_{mn} ({\bf x})$ enables defining pure elastic strain and stress following the convention \cite{Mura}
\begin{equation}
    \epsilon _{ij}({\bf x}) = \epsilon ^{(tot)} _{ij}({\bf x}) - \omega _{ij}({\bf x}).
    \label{elastic_strain}
\end{equation}
and 
\begin{eqnarray}
    \sigma_{ij}({\bf x})   &=& C_{ijkl}\epsilon_{kl}({\bf x}) 
                            = \sigma^{(tot)}_{ij}({\bf x})-\Pi_{ij}({\bf x})\nonumber \\ 
                            &=& \sigma^{(tot)}_{ij}({\bf x})-C_{ijkl}\omega _{kl}({\bf x}),
    \label{elastic_stress}
\end{eqnarray}
where $\sigma ^{(tot)}_{ij}({\bf x}) =C_{ijkl}\epsilon ^{(tot)} _{kl}({\bf x})$. The total strain in \eqref{elastic_strain} is taken as a symmetrised derivative of atomic displacements \cite{LandauElasticity}
\begin{equation}
    \epsilon ^{(tot)} _{kl}({\bf x})    = {1\over 2}\left({\partial u_k({\bf x})\over \partial x_l} 
                                        + {\partial u_l({\bf x}) \over \partial x_k} \right),
    \label{total_strain_definition}
\end{equation}
and hence equations \eqref{elastic_strain} and \eqref{elastic_stress} define the purely elastic components of spatially varying strain and stress.  
Equations \eqref{elastic_strain} and \eqref{elastic_stress} show that at the macroscopic level, elastic strain and stress fields are defined taking the notion of a material containing a certain local concentration of defects, corresponding to eigenstrain $\omega_{ij}({\bf x})$, as a reference non-distorted configuration.  

Defects can be produced not only by irradiation but also by processing, for example for the purpose of creating a microstructurally complex material. A complex microstructure can be viewed as a highly imperfect material, characterised by a high density of regions where local atomic structure deviates significantly from ideal crystalline order. An example of a microstructurally complex material is steel, where strongly fluctuating plastic strains and deformations are introduced during manufacturing by alloying and through thermo-mechanical treatments.

In the absence of conventional macroscopic body forces, for example gravity or thermal expansion \cite{LandauElasticity}, the condition of mechanical equilibrium \cite{Sutton2020} $\partial \sigma _{ij}({\bf x})/\partial x_j=0$, expressed in terms of elastic stress \eqref{elastic_stress}, has the form 
\begin{equation}\label{elastic_equilibrium}
    C_{ijkl}{\partial \epsilon ^{(tot)}_{kl}({\bf x})\over \partial x_j} - C_{ipmn}{\partial \omega_{mn}({\bf x})\over \partial x_p} = 0.
\end{equation}
Interpreting this as a condition of equilibrium for the {\it total} stress, namely $\partial \sigma _{ij}^{(tot)}({\bf x})/\partial x_j+f_i({\bf x})=0$, from equation \eqref{elastic_equilibrium} we see that the spatially varying eigenstrain \eqref{volume_tensor_density} can be interpreted as a source of an effective body force \cite{NF2018}
\begin{equation}\label{effective_body_force}
    f_i({\bf x}) = -C_{ipmn}{\partial \omega_{mn}({\bf x}) \over \partial x_p} = -{\partial \over \partial x_p}\Pi _{ip}({\bf x}).
\end{equation}
This interpretation takes the ideal crystalline state of the material as a reference. 
In this picture, radiation swelling originates from the internal body forces generated by the distribution  of defects defined by equation \eqref{effective_body_force}. These forces act on the ideal atomic structure of the material, causing it to deform.   

A convenient feature of pure elastic strain \eqref{elastic_strain} is that the integral of the trace of the total strain over the volume $V$ of the region occupied by the material, subject to traction-free boundary conditions, equals the sum of relaxation volumes of all the defects in it \cite{NF2018,Leibfried1978,Kossevich}
\begin{equation}\label{total_volume}
    \int_V \epsilon_{ii}^{(tot)}({\bf x}) \dint{^3x} = \int_V \omega_{ii}({\bf x}) \dint{^3x}. 
\end{equation}
The trace of pure elastic strain vanishes after integration over the same region
\begin{equation}\label{trace_elastic_strain}
    \int_V \epsilon_{ii}({\bf x}) \dint{^3x} = 0. 
\end{equation}
This condition shows that the trace of pure elastic strain $\epsilon _{ii}({\bf x})$ either vanishes identically everywhere inside the material, or changes sign, varying from compression to tension. A detailed proof, given in Appendix~\ref{sec:A:relaxation}, shows that $\epsilon _{ij}({\bf x})$ in fact satisfies an even stronger condition  
\begin{equation}
    \int_V \epsilon _{ij}({\bf x}) \dint{^3x} = 0.
    \label{volume_average_elastic_strain}
\end{equation}
The latter is particularly significant since it shows that it is not only the elastic strain that averages to zero over the volume of the material. The elastic stress, which is a linear combination of elements of the elastic strain tensor, must also vanish after the integration over the volume of the material, provided that the traction-free boundary conditions are satisfied; namely
\begin{equation}
    \int_V \sigma _{ij}({\bf x}) \dint{^3x} = 0. 
    \label{volume_average_elastic_stress}
\end{equation}
Below, we show that this condition of vanishing mean stress proves highly significant in applications.

It is the total {\it local} strain $\epsilon _{ij} ^{(tot)}({\bf x})$ that is observed in X-ray diffraction experiments. For example, a negative value of total strain, signifying lattice contraction, is observed in an elastically unstrained material containing high spatially homogeneous concentration of vacancies \cite{Hertz1973}. On the other hand, large positive strain associated with the accumulation of self-interstitial atom defects is observed in thin near-surface layers of tungsten implanted with self-ions to a relatively low dose \cite{Mason2020}. The integral of the trace  of the total strain $\epsilon _{ii}^{(tot)} ({\bf x})$ over the geometric volume of the material determines the magnitude of dimensional changes and the amount of volumetric swelling occurring following exposure to radiation, see Appendix~\ref{sec:A:volumechange} for more detail. 

We highlight the macroscopic nature of distinction between the total and pure elastic components of strain. At a microscopic level, where we are interested in the local deformation of the lattice in the vicinity of a particular defect, there is no need to separate the total and elastic components of strain. However, at the macroscale, where the density of relaxation volumes of defects \eqref{volume_tensor_density} becomes a meaningful macroscopic quantity, separating the total and elastic components of strain is fully warranted and in fact necessary.   

In what follows, we explore several examples illustrating how elastic strain and stress, \eqref{elastic_strain} and \eqref{elastic_stress}, vary in reactor components exposed to irradiation. We find elastic strain and stress fields in a rectangular-shape component where the distribution of defects is either spatially homogeneous or spatially heterogeneous. This geometry is also commonly used in ion implantation experiments. 

We also investigate the analytically tractable cases of components with cylindrical and spherical geometries exposed to symmetric sources of radiation. Pressurised cylindrical pipes are used in neutron irradiation tests probing how materials respond to exposure to neutrons in the presence of external stress, generated by the internal gas pressure in a pipe. A spherical shell provides a convenient generic case for exploring radiation effects in materials since the displacement, strain, and stress fields turn out to depend only on one, radial, variable. This enables finding exact analytical solutions suitable for comparison with numerical FEM analysis. Abaqus 2021 computer program was used for this numerical work.  The case studies given below enable exploring the effect of spatial variation of defect densities, stemming from the variation of neutron flux, on the magnitude of elastic strain and stress developing in reactor components following their exposure to irradiation.

We find that in the absence of constraints imposed by other elements of the reactor structure, a spatially homogeneous distribution of defects gives rise to volumetric swelling but produces no elastic stress in a component. On the other hand, a strongly spatially heterogeneous distribution of defects, stemming from a strongly spatially varying radiation exposure, may give rise to stress concentrations of significant magnitude, of both compressive and tensile character. 

\section{Elastic field of point defects}

In this section, we provide the general formulae describing the elastic field of point defects produced in a material by exposure to irradiation. We assume the validity of isotropic elasticity approximation, and also that the defects adopt statistically random orientations \cite{NF2018}, resulting is that the individual relaxation volume tensors acquire the form
\begin{equation}
    \Omega ^{(a)}_{mn} = {1\over 3}\Omega ^{(a)} \delta _{mn}, \label{volume_isotropic}
\end{equation}
where $\delta_{mn}$ is the Dirac delta-symbol, satisfying the condition $\delta_{nn}=3$. 
As a result, the spatially-dependent eigenstrain averaged over all orientations is also diagonal with respect to indices $m$ and $n$, namely 
\begin{equation}\label{eigenstrain}
    \epsilon ^*_{mn}({\bf x}) = \omega _{mn}({\bf x}) = {1\over 3} \omega ({\bf x})\delta _{mn},
\end{equation}
where $\omega ({\bf x})$ is the density of relaxation volumes of defects \cite{NF2018}. Multiplying (\ref{eigenstrain}) by the tensor of elastic constants of the material, which in the isotropic approximation has the form \cite{Mura,Sutton2020}
\begin{equation}\label{elastic_constants}
    C_{ijkl} = \mu {2\nu \over 1-2\nu} \delta_{ij} \delta_{kl} 
             + \mu \left( \delta_{ik}\delta_{jl} + \delta_{il}\delta_{jk} \right), 
\end{equation}
where $\nu $ is the Poisson ratio and $\mu $ is the shear modulus, we find the eigenstress \cite{Mura} generated by the defects
\begin{equation}\label{eigenstress}
    \sigma ^*_{ij}({\bf x}) = C_{ijkl}\epsilon ^*_{kl}({\bf x}) = B\omega ({\bf x})\delta _{ij}. 
\end{equation}
In the above equation, $B$ is the bulk modulus of the material \cite{CaiNix2016,NF2018}
\begin{equation}
    B={2\mu \over 3} \left( {1 + \nu \over 1-2 \nu} \right).
\end{equation}
The field of displacements generated by an isotropic point defect with relaxation volume $\Omega$ situated at point ${\bf x}'$ is \cite{Kossevich,Eshelby1956}
%
%
\begin{align}
    u_i({\bf x})    &= \phantom{-}\frac{\Omega}{12\pi} \left(\frac{1+\nu}{1-\nu}\right) 
                            \frac{x_i-x_i'}{|{\bf x}-{\bf x}'|^3} \nonumber \\
                    &= -\frac{\Omega}{12\pi} \left(\frac{1+\nu}{1-\nu}\right)
                            \frac{\partial}{\partial x_i} \frac{1}{|{\bf x}-{\bf x}'|},
    \label{displacements_point_defect}
\end{align}
which follows from equation~\eqref{eq:displacement_eigenstrain} taken in the isotropic elasticity approximation, with the density of relaxation volume tensors given by $\omega_{mn}({\bf x}) = \frac{1}{3}\Omega \delta_{mn} \delta({\bf x} - {\bf x}')$.

Kossevich \cite{Kossevich} gives a detailed analysis of the above formula for the field of displacements produced by an isotropic point defect and discusses the range of its validity. In particular,  \cite{Kossevich} highlights the fact that the field of displacements given by equation \eqref{displacements_point_defect} does not take into account the boundary conditions at surfaces, and hence an attempt to evaluate the relaxation volume of a defect directly from \eqref{displacements_point_defect}, neglecting the boundary conditions, produces an unexpectedly wrong result.  

Similarly, the field of displacements generated directly by a distribution of point defects and not including the effect of boundary conditions, is
%
%
\begin{equation}\label{displacements}
    u_i({\bf x}) =  \frac{1}{12\pi} \left(\frac{1+\nu}{1-\nu}\right)
                    \int_V \omega ({\bf x}') \frac{x_i - x_i'}{|{\bf x}-{\bf x}'|^3} \,\dint{^3x'}.
\end{equation}
The total strain can now be computed using equation \eqref{total_strain_definition}.
Comparing equations \eqref{displacements} and \eqref{total_strain_definition}, 
we see that the trace of the total strain tensor $\epsilon ^{(tot)} _{ii}({\bf x})$, computed directly from  \eqref{displacements}, equals
\begin{equation}\label{eq_trace_epsilon}
    \epsilon^{(tot)}_{ii}({\bf x}) = \frac{1}{3} \left(\frac{1+\nu}{1-\nu}\right) \omega ({\bf x}).
\end{equation}
The fact that the right-hand-side of the above equation is not equal to $\omega ({\bf x})$ illustrates the fact that the total change of volume of the material due to the presence of defects is only partially associated with the direct sources of strain, i.e. the local deformations of the lattice produced by the defects, and that partially it arises from the elastic fields formed as a result of application of boundary conditions \cite{Leibfried1978,Kossevich}.

A common example of eigenstrain is thermal strain. Equation (1.1) by Mura \cite{Mura} states that in the isotropic approximation $\epsilon^*_{ij}({\bf x})=\delta_{ij}\alpha T({\bf x})$, where $\alpha$ is the linear thermal expansion coefficient and $T({\bf x})$ is the temperature, provided that the body is in its reference state at $T=0$. Applying the condition of mechanical equilibrium, we see that the above thermal strain produces an effective body force  \cite{LandauElasticity}
\begin{equation}\label{eq:f_thermal}
    f_i=-3B\alpha\frac{\partial T}{\partial x_i}.
\end{equation}
Inserting \eqref{volume_isotropic} and \eqref{elastic_constants} into \eqref{effective_body_force} and comparing it with \eqref{eq:f_thermal} shows a way of evaluating stress and strain fields through the use of a fictitious thermal expansion term in FEM, defined by $\alpha=1/3$ and $T({\bf x})=\omega({\bf x})$. A similar approach was followed by Nguyen {\it et al.} \cite{Kurtz2017} and Leide {\it et al.} \cite{Leide2020}.

Concluding this section, we note that the isotropic formula for the volume tensor of a defect (\ref{volume_isotropic}) used in a phenomenological treatment of swelling \cite{Kurtz2017,Hofmann2017,NF2018,Leide2020} does not represent a generally valid approximation. There are documented cases showing that externally applied stress, or the stress developing in a material under irradiation, can polarise defects and generate eigenstrain not diagonal with respect to its Cartesian components \cite{Chen2010,Boleininger2019,Mason2020,Lai2020}. In this study, we do not explore the effects of elastic polarization of defects, but note that equations (\ref{elastic_strain}) --
(\ref{effective_body_force}) enable the evaluation of elastic deformation, strain and stress for an arbitrary anisotropic distribution of eigenstrain produced by the defects generated in materials by irradiation.

\section{\label{sec:rectangle}Elastic stress in an ion-irradiated thin film} 
\begin{figure}[t]
  \includegraphics[width=0.45\textwidth]{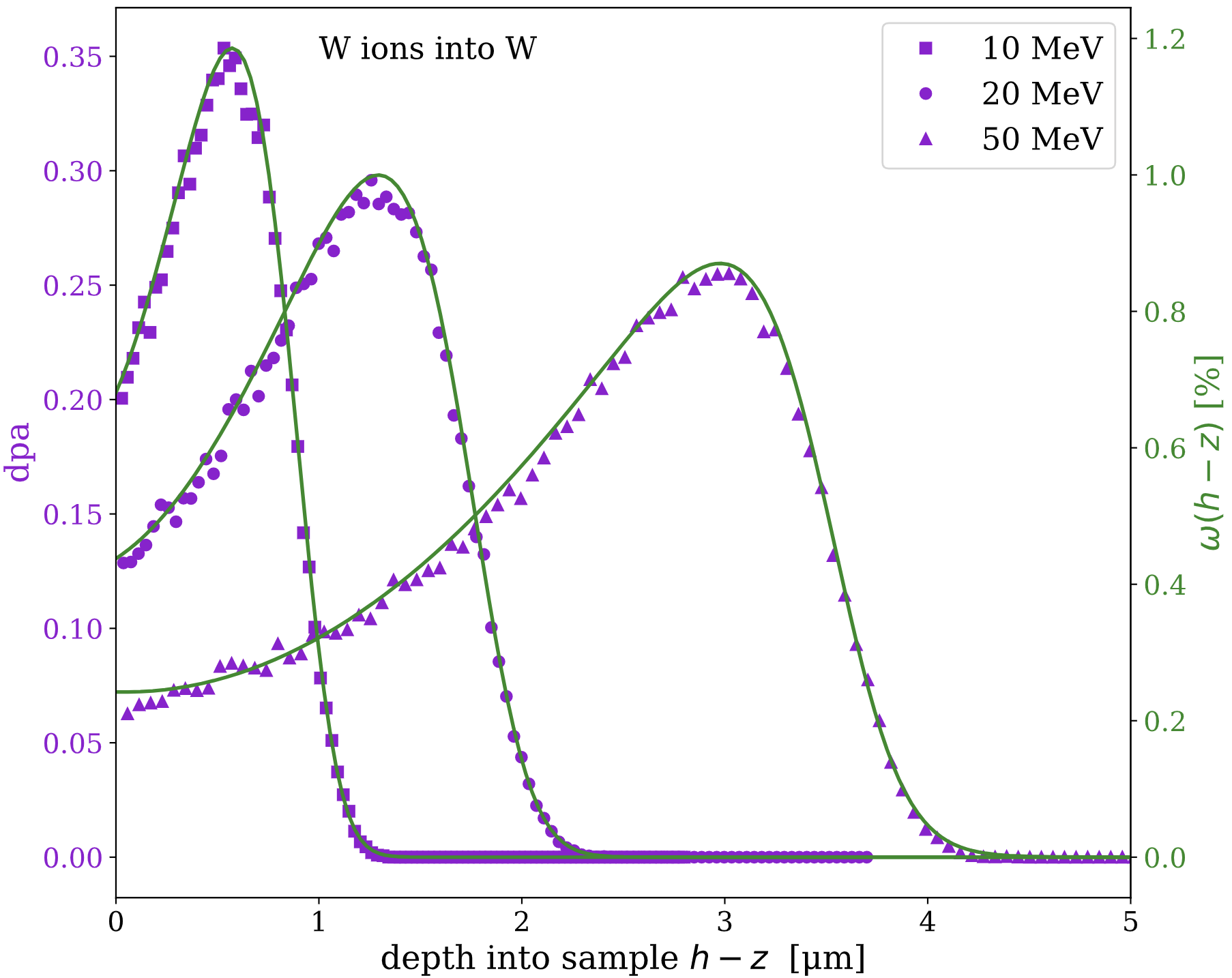}%
    \caption{Displacement damage profile caused by W ions implanted into W, calculated using SRIM and corresponding to an incident ion fluence of $10^{14}$ ions/cm$^2$. The data are fitted using equation \eqref{eq:fit_dpa}. The relaxation volume density used in simulations of elastic stress and strain fields produced by ion irradiation is assumed to follow the same spatial profile, and is scaled such that the relaxation volume density curve for the 20 MeV case peaks at 1\%. For the definition of depth and the coordinate system used in simulations see Sec. \ref{sec:rectangle}. }
    \label{fig:dpa_profile_film}
\end{figure}

Ion irradiation is used as a cost- and time-effective surrogate for neutron irradiation when studying changes in materials properties in a reactor environment. High energy ions have a shorter penetration depth compared to neutrons, and the damaged layer in an ion-irradiated sample is typically only a few microns in thickness. The implantation profile depends on the energy of the ions and can be estimated using the SRIM~\cite{ZIEGLER20101818,srim2008} Monte Carlo simulation code. Fig. \ref{fig:dpa_profile_film} shows the displacement damage profile (in displacements per atom, or dpa) for W ions implanted into W. The data points shown were obtained from SRIM via the so-called Kinchin-Pease ``quick'' calculation (as recommended in~\cite{STOLLER201375,Agarwal2021}). For ion energies of 10, 20 and 50~MeV, the implantation of 10000 W ions were simulated and the resulting vacancies/ion as a function of depth into the sample were used to calculate the dpa for an incident ion fluence of \(10^{14}\)~ions/cm\(^2\), which is typical of low-dpa ion-irradiation experiments (see, for example, table S1 in the supplemental material of~\cite{Mason2020}). Note that, for these SRIM simulations, the ASTM standard threshold displacement energy of 90~eV (see, e.g.,~\cite{MACFARLANE20102739}, table II) was assumed for W.

Because of the short ion implantation range, samples for ion-irradiation studies are commonly chosen as thin foils, with micro-mechanical tests subsequently applied to characterize the mechanical properties of the highly stressed damaged layer. For this purpose, it is desirable to be able to fully quantify the relation between the residual stress, or eigenstress, and the elastic fields in the sample. In the limit where the foil thickness is small relative to its width, we are able to find analytical solutions for the elastic stress and strain fields resulting from the accumulation of defects in the foil. 

The general solution for the field of displacements is obtained by solving for the condition of mechanical equilibrium \eqref{elastic_equilibrium}, which is a linear inhomogeneous partial differential equation. In what follows, we solve the problem using a standard approach by first finding a particular solution of equation \eqref{displacements}, and then by finding a solution to the homogeneous problem, which follows from the condition of mechanical equilibrium \eqref{elastic_equilibrium} for $\omega_{mn}({\bf x}) = 0$. The full solution is given by the sum of the particular and homogeneous solutions, where the unknown constants entering the homogeneous solution are identified from the boundary conditions.

Consider a foil with its planar surface parallel to the $xy$-plane, irradiated by ions coming from the $z$-direction, producing a relaxation volume density profile  $\omega(z)$. As the foil width in $x$- and $y$-directions is considered to be large relative to its thickness, and the foil is attached to a substrate such that it is unable to buckle, we assume that the strain and stress fields are independent of $x$ and $y$ coordinates. Taking the ensemble-averaged density of relaxation volume tensors (\ref{volume_tensor_density_ensemble}) as diagonal with respect to the Cartesian indices \cite{NF2018}, we write
\begin{equation}
    \omega_{mn}({\bf x}) = \frac{1}{3} \omega(z) \delta_{mn}.
\end{equation}
The field of displacements is defined on the interval $0 \leq z \leq h$, where $h$ is the thickness of the foil. The particular solution, excluding the effect of boundary conditions, follows immediately from \eqref{eq_trace_epsilon} as
\begin{equation}\label{eq:uzz_film}
    \frac{\partial}{\partial z} u_z(z) = \frac{1}{3} \left(\frac{1+\nu}{1-\nu}\right) \omega(z),
\end{equation}
where we noted that the in-plane displacements vanish in the thin film limit, as can be confirmed by explicitly integrating \eqref{displacements}. The field of displacements can now be found by integrating \eqref{eq:uzz_film} with respect to $z$, namely 
\begin{equation}\label{eq:uz_film}
    u_z(z) = \frac{1}{3} \left(\frac{1+\nu}{1-\nu}\right) \int_0^z \omega(\zeta ) \dint{\zeta }.
\end{equation}

We now need to find a solution to the homogeneous problem. Since the strain field does not depend on $x$ or $y$, the field of displacements has the form
\begin{equation}\label{eq:uz_film_hmg}
    {\bf u }({\bf x }) = c x\,  {\bf e}_{x} + c y\, {\bf e}_y + u_z(z)\, {\bf e}_z,
\end{equation}
where ${\bf e}_x$, ${\bf e}_y$ and ${\bf e}_z$ are the Cartesian unit vectors, and $c$ is a constant. Substituting \eqref{eq:uz_film_hmg} into the homogeneous condition of mechanical equilibrium, we arrive at an ordinary differential equation 
\begin{equation}
    \frac{\mathrm{d}^2 u_z(z)}{\mathrm{d}z^2} = 0.
\end{equation}
This equation has general solutions of the form
\begin{equation}\label{eq:uz_film_sol}
    u_z(z) = a + b z,
\end{equation}
where we are free to set $a\equiv0$ as uniform displacements do not affect elastic fields.

A general solution for ${\bf u }({\bf x})$, valid in the interval $0 \leq z \leq h$, is given by the sum of the homogeneous solution \eqref{eq:uz_film_sol} and the particular solution \eqref{eq:uz_film}, namely
\begin{equation}\label{eq:u_general_bc}
\begin{aligned}
    u_x(x) &= c x \\
    u_y(y) &= c y \\
    u_z(z) &= b z + \frac{1}{3} \left(\frac{1+\nu}{1-\nu}\right) \int_0^z \omega(\zeta ) \dint{\zeta }.
\end{aligned}
\end{equation}
Combining equations \eqref{elastic_strain}, \eqref{elastic_stress}, \eqref{eigenstress} with the above expression (\ref{eq:u_general_bc}), we can express the non-vanishing components of the elastic stress tensor as
%
%
%
\begin{equation*} 
\begin{aligned}
\sigma_{xx}(z) = \sigma_{yy}(z) &= \frac{2\mu(c+b\nu)}{1-2\nu} 
                            - \frac{2\mu}{3}\left(\frac{1+\nu}{1-\nu}\right) \omega(z), \\
\sigma_{zz}(z) &= \frac{2\mu b(1-\nu)}{1-2\nu} + \frac{4\mu c \nu}{1-2\nu}.
\end{aligned}
\end{equation*}
All the other components of tensor $\sigma _{ij}(z)$ are equal to zero.  

Constants $b$ and $c$ depend on the conditions at the top and bottom surfaces of the foil. In the context of an ion-irradiation experiment, the surface of the foil at $z=h$, facing the ion flux, is free of tractions, therefore the first condition is $\sigma_{zz} (h) = 0$. For the second condition, we consider first the case where the foil is able to relax freely in the $xy$-plane. While it is not possible to derive a solution where the foil sides are completely free of tractions, as that would require buckling of the foil, we can weaken the condition and instead assert that the average traction at the sides of the foil vanishes, which corresponds to a sliding condition. For example, the average traction over the foil boundaries at $x\rightarrow \pm \infty$ is
\begin{equation}\label{eq:foilside}
\frac{1}{h}\int_0^h t_i \dint{z} = \frac{1}{h}\int_0^h \sigma_{ij} n_j \dint{z} = \frac{1}{h}\int_0^h  \sigma_{xx}(z)\dint{z}=0,
\end{equation}
where $\vec{n} = {\bf e}_x$ is the outward surface normal vector to the side of the foil. Solving for $b$ and $c$ such that conditions $\sigma_{zz} (h)= 0$ and \eqref{eq:foilside} are met, we arrive at explicit analytical expressions for the field of displacements
\begin{equation}\label{eq:u_free_bc}
\begin{aligned}
    u_x(x) &= \frac{1}{3} \wmean x \\
    u_y(y) &= \frac{1}{3} \wmean y \\
    u_z(z) &= -\frac{2\nu}{3(1-\nu)} \wmean z 
                    + \frac{1}{3} \left(\frac{1+\nu}{1-\nu}\right) \int_0^z \omega(\zeta ) \dint{\zeta },
\end{aligned}
\end{equation}
where $\wmean$ is the mean value of function $\omega(z)$ over the thickness of the foil
\begin{equation}
  \wmean = \frac{1}{h} \int_0^h \omega(\zeta )\dint{\zeta }.
\end{equation}
The only elastic stress components that do not vanish are the in-plane stresses, which equal
\begin{equation}\label{eq:sigma_free_bc}
    \sigma_{xx}(z) = \sigma_{yy}(z) = -\frac{2\mu}{3} \left(\frac{1+\nu}{1-\nu}\right)
                                            \left[\omega(z) - \wmean\right].
\end{equation}
\begin{figure}
\subfloat[\label{fig:film_p_u_z}]{%
  \includegraphics[width=0.4\textwidth]{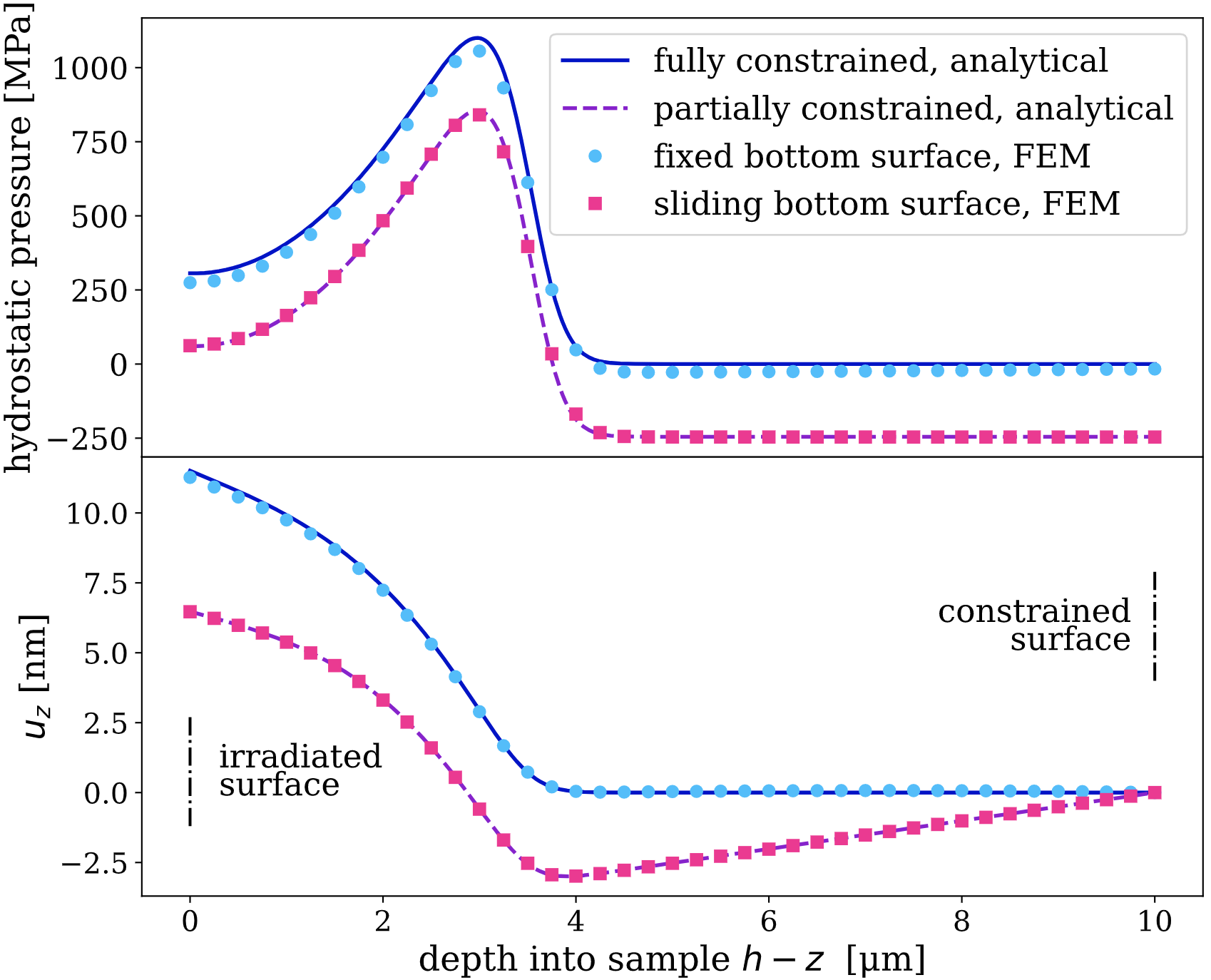}%
}\hfill
\subfloat[\label{fig:film_Sxx_FEM}]{%
  \includegraphics[width=0.4\textwidth]{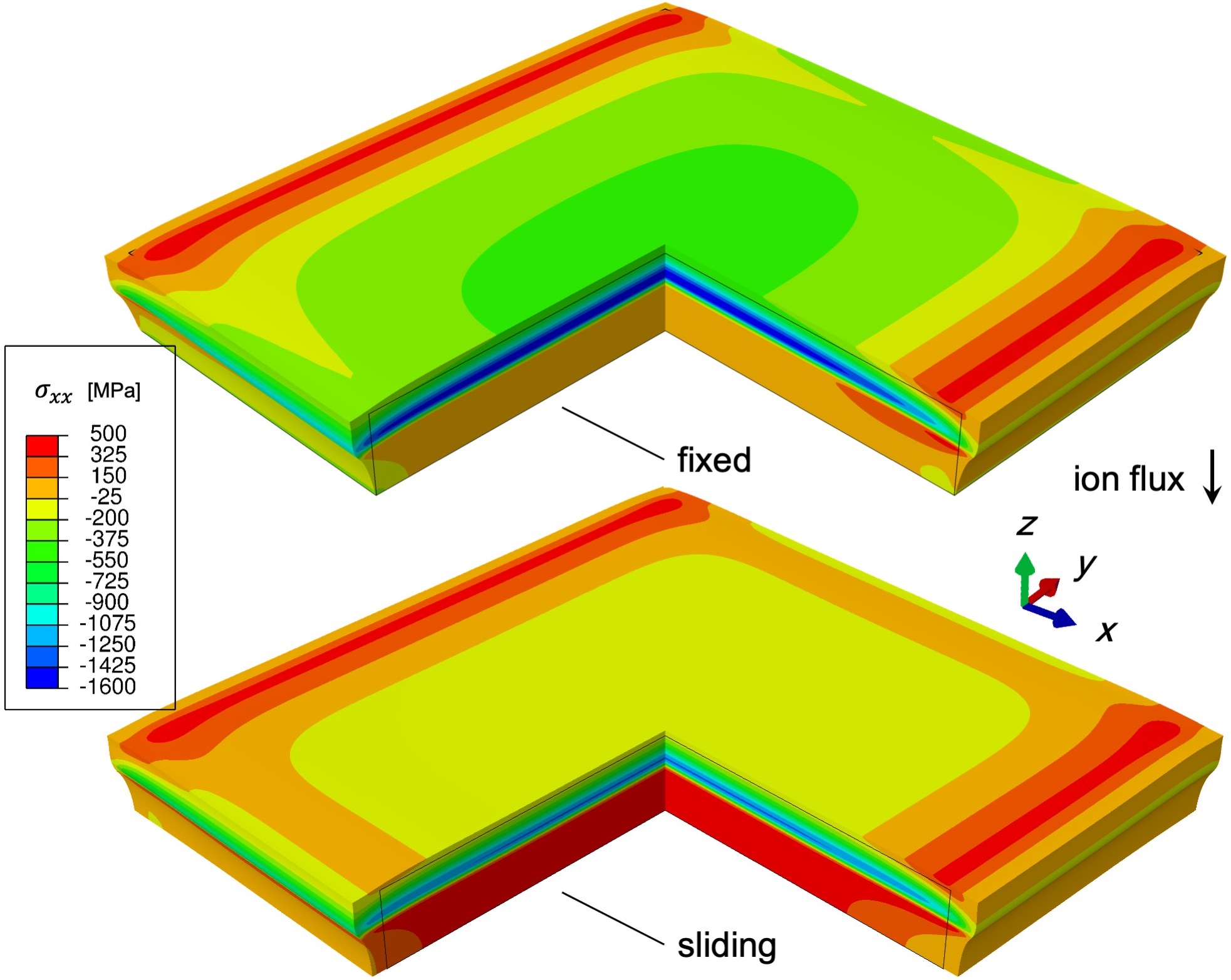}%
}\hfill
\subfloat[\label{fig:film_p_FEM}]{%

  \includegraphics[width=0.4\textwidth]{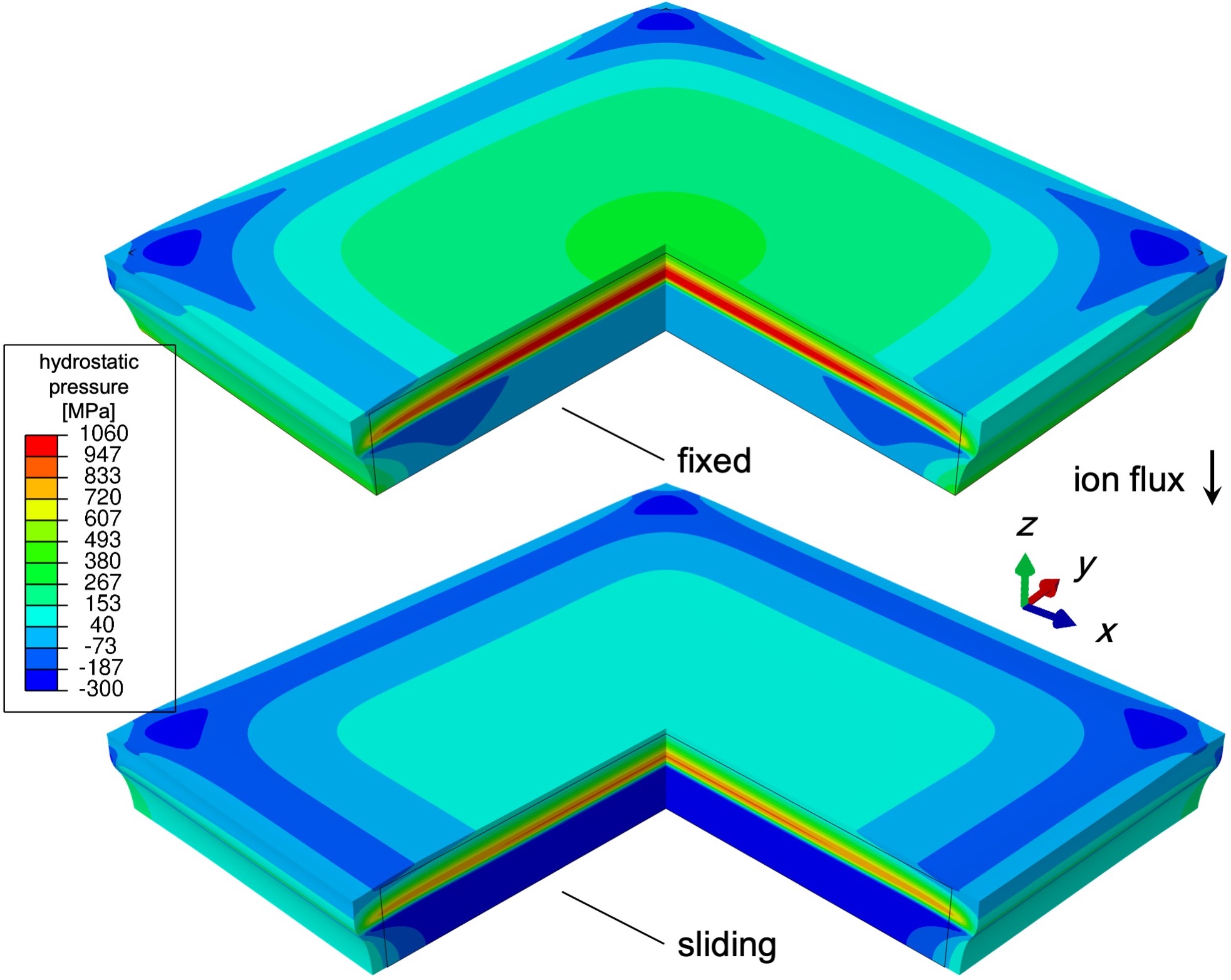}%
  }\hfill
    \caption{Swelling in an ion-irradiated thin foil only extends over a few \textmu m. The presence of the unirradiated material results in a high compressive in-plane stress in the irradiated region. Depending on the boundary conditions at the bottom of the foil, the unirradiated material is either unstressed or is under tension. Panel (a) shows that, by preventing or allowing for motion of the bottom surface in $x$ and $y$, the FEM solution approaches the two analytical solutions derived in this section. Panels (b) and (c) show the contour plots of $\sigma_{xx}({\bf x})$ and the hydrostatic pressure $p({\bf x})=-{1\over 3}\sigma _{ii}({\bf x})$ in the sectioned view of the film. The initial profile of the foil before deformation is also outlined. The density of relaxation volumes matches the profile of 50 MeV W ion implantation shown in Fig. \ref{fig:dpa_profile_film}.}
    \label{fig:film_plots}
\end{figure}

Let us now consider the case where the sides of the foil are fixed such that no in-plane displacements occur. The displacement field of the constrained foil $u^C_i({\bf x})$ with the condition $u^C_x = u^C_y = 0$ is obtained from the sliding foil solution \eqref{eq:u_free_bc} by setting $\wmean = 0$, irrespective of its actual value, as then the in-plane displacements vanish. The field of vertical displacements is now given by
\begin{equation}\label{eq:u_fixed_bc}
    u_z^C(z) = \frac{1}{3} \left(\frac{1+\nu}{1-\nu}\right) \int_0^z \omega(\zeta ) \dint{\zeta }.
\end{equation}
The in-plane components of elastic stress are 
\begin{equation}\label{eq:sigma_fixed_bc}
    \sigma_{xx}^C(z) = \sigma_{yy}^C(z) = -\frac{2\mu}{3} \left(\frac{1+\nu}{1-\nu}\right) \omega(z).
\end{equation}
Incidentally, the same solution is found by taking the limit $h \rightarrow \infty$ in the solution for a sliding foil  \eqref{eq:u_free_bc}, provided that the damage profile is localised such that ${\lim_{h\rightarrow\infty}\overline{\omega} = 0}$. This demonstrates that the in-plane stresses in the solution for a sliding foil approach those of the constrained foil as the foil thickness increases.

Solutions found for the constrained and unconstrained cases show that swelling induces stress only in the plane of the foil, and there is no stress in the direction normal to the foil surface, either at the surface or in the bulk of the foil. The in-plane stress can be fairly high, for example swelling of \SI{1}{\percent} in a constrained tungsten foil produces an in-plane stress approaching \SI{-1.9}{\GPa}.

To verify the analytical solutions given above, we compare them to numerical FEM simulations. We considered a tungsten foil ($\mu=\SI{160}{\GPa}$ and $\nu=0.28$) with dimensions of \SI{80}{\micro\meter} in the $xy$-plane and height of $h=\SI{10}{\micro\meter}$ along $z$. Two types of boundary conditions were considered at the bottom surface of the foil interfacing the substrate: The foil is either completely constrained ($u_x(0)=u_y(0)=u_z(0)=0$) or its displacements are constrained only in the $z$ direction ($u_z(0)=0$). The FEM solutions involved approximately $1.20\times10^6$ hexahedral linear elements. The free surface of the foil before irradiation is at $z=h$, and the assumed profile $\omega (z)$ is shown in Fig. \ref{fig:dpa_profile_film} as a function of the distance from the top surface, or $h-z$. Fig. \ref{fig:film_plots}, where the density of relaxation volumes corresponding to 20 MeV ion irradiation was used as input, shows that the two analytical solutions derived in this section approach the two types of boundary conditions considered for the FEM numerical solutions. 

\begin{figure}[t]
  \includegraphics[width=0.48\textwidth]{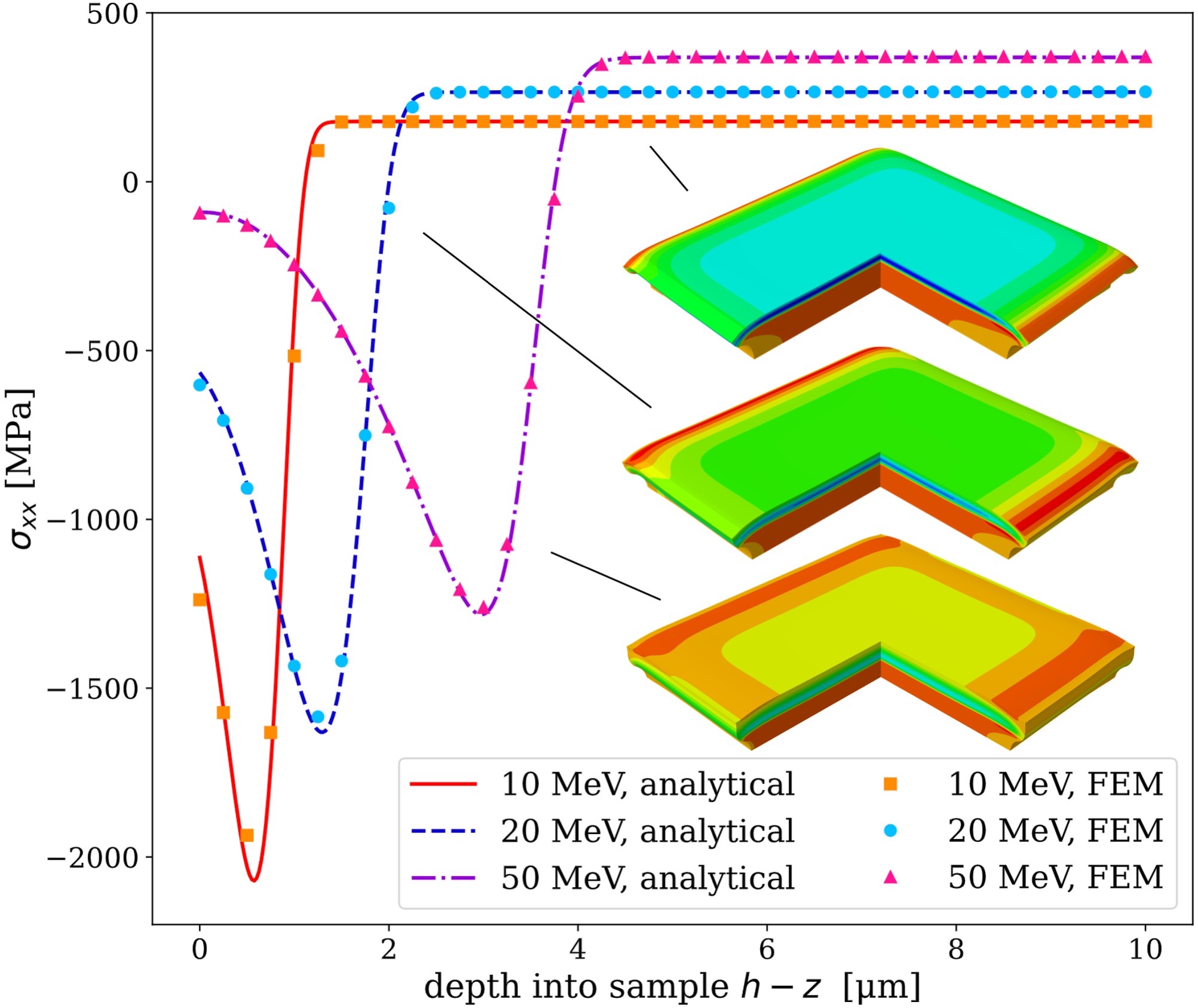}%
    \caption{$\sigma_{xx}(z)=\sigma_{yy}(z)$ at the centre of a rectangular foil exposed to ion irradiation and plotted as a function of depth $h-z$. The analytical solutions and FEM numerical solutions are in full agreement and show that the irradiation-induced stresses follow the variation of the density of relaxation volumes (cf. Fig. \ref{fig:dpa_profile_film}), where the profiles generated by ions with higher energy correspond to lower compressive stress in the irradiated region and higher tensile stress in the unirradiated region underneath. The colour scale is common for all the three plots and extends between -2000 MPa (blue) and 600 MPa (red). The bottom surface of the foil is free to expand in the in-plane directions.}
    \label{fig:Sxx_film_E}
\end{figure}

It is desirable to quantify the magnitude of elastic stress that develops in a foil following ion irradiation. Indeed, the ion irradiation data can help extrapolate experimental observations to the neutron irradiated case, where the distribution of relaxation volume density $\omega ({\bf x})$ varies on the scale of tens of centimetres \cite{Gilbert2013}. Fig. \ref{fig:Sxx_film_E} shows how the in-plane components of stress vary depending on the energy of the incident ions, which produce damage at greater depth with higher energy. The boundary condition at the bottom surface is taken to be the sliding one, allowing for expansion of the film in the $xy$-plane. The $\sigma_{zz}$ component of stress was found to be vanishingly small in all cases (lower than \SI{1}{\MPa}), confirming the results of analytical treatment. The in-plane compressive stress components are significant, and exceed \SI{-1}{\GPa} in the irradiated layer even if the foil is able to relax and expand sideways. 

The development of compressive or dilatational stresses in the GPa range has been observed in a variety of materials exposed to ion irradiation, including ceramic oxides (\SI{-2.7}{\GPa} in Gd$_2$TiZrO$_7$ at 0.2 dpa) \cite{Sattonnay2008},  SiC and SiO$_2$ \cite{Harbsmeier1998}, Cr  \cite{Misra1998}, Au \cite{Eren2014} and W \cite{Mason2020} thin films. Davis \cite{Davis1993} estimated the compressive stress in thin films growing on a substrate, assuming that a steady state is attained when the rate of implantation of new ions is balanced by the rate with which stress is relaxed by atoms escaping to the surface. Wolfer and Garner proposed instead that the compressive stress depends on the competition between swelling and creep \cite{Wolfer1979}. These examples show that taking into account the effects of stress relaxation,\textemdash whether by thermal or irradiation creep,\textemdash is required before comparing theoretical predictions with experimental results. 

A detailed study of strain and stress in ion-irradiated tungsten \cite{Mason2020,Mason2021} shows that the compressive stress in the ion-irradiated layer initially reaches $\sim320$ MPa \footnote{$320$ MPa is approximately equivalent to $\sim0.002$ eV/\AA$^3$ in atomic units. This conversion can be performed using a convenient rule where 1 eV/\AA$^3\;$ equals 160.2176621 GPa.} at 0.3 dpa before changing sign and equilibrating, at an approximately similar level in magnitude but {\it dilatational} in character, in the high dose limit. This finding agrees with the earlier observations by Misra {\it et al.} \cite{Misra1998} and illustrates the highly non-linear nature of the stress relaxation phenomenon. It appears that the radiation-driven relaxation still does not fully eliminate the stress produced by the defects themselves, resulting in the residual stress in the several hundred MPa range \cite{Mason2020,Henry2019}. 

\section{\label{sec:cylinder}Irradiation-induced elastic stress and strain in a cylindrical tube} 
In this section we investigate the distribution of elastic strain and stress in a sample of cylindrical geometry. Cylindrical tubes are often used in the context of neutron irradiation experiments not only because of the natural similarity with the geometry of fission fuel cladding but also because a sample of cylindrical geometry offers a convenient way of testing the combined effect of externally applied stress and irradiation. In practice, the external stress is produced by the pressure of gas inside the cylinder, whereas the exposure to irradiation is often external. 

Consider a cylindrical tube with the inner radius $\rho_1$ and outer radius $\rho_2$, and define a cylindrical coordinate system $(\rho, \varphi, z)$ coaxial with the tube. Pressure of gas $P$ applied to the inner surface is assumed to be constant, and the density of relaxation volumes tensor is a function of only the radial coordinate $\rho$, namely
\begin{equation}\label{eq_omega_cylinder}
    \omega_{mn}({\bf{x}})=\frac{\omega(\rho )}{3}\delta_{mn}.
\end{equation}
Since neither $P$ nor $\omega (\rho) $ depend on $\varphi$ or $z$, the field of displacements has the form
\begin{equation}\label{eq:displ_cyl_general}
    \textbf{u}({\bf{x}})=u_\rho(\rho){\bm{e_\rho}}+c z\, {\bm{e}_z},
\end{equation}
where $c$ is a constant and ${\bm{e_\rho}}$ and ${\bm{e}_z}$ are the unit vectors in the radial and $z$ directions, respectively. The choice of the above simple form of the field of displacements stems from the fact that $u_\varphi=0$ from symmetry, and the axial strain must be translationally invariant with respect to $z$, implying that  $\partial u_z/\partial z=c$. For a pipe of finite length, this translational invariance only applies to the central portion of the pipe, far from its two extremities. Below, we explore two different boundary conditions along $z$, namely that the pipe is either fully unconstrained or fully constrained. We find that the irradiation-induced swelling produces different patterns of stress in these two limits.

If the tube is constrained in the $z$ direction, then $c=0$. If the distribution of  ${\omega_{mn}({\bf{x}})}$ is entirely spatially uniform, the components of stress can be found by solving the homogeneous equation ${\partial\sigma_{ij}/\partial x_j=0}$. Evaluating the divergence of stress in the cylindrical system of coordinates in the isotropic elasticity approximation, the field of radial displacements can be found by solving the ordinary differential equation
\begin{equation}
    \rho\frac{\text{d}^2u_\rho}{\text{d}\rho^2}+\frac{\text{d}u_\rho}{\text{d}\rho}-\frac{u_\rho}{\rho}=0.
\end{equation}
A general solution of this homogeneous equation has the form \cite{LandauElasticity} 
\begin{equation}\label{homogeneous_cylindrical}
u_\rho(\rho) = a\rho + \frac{b}{\rho}.
\end{equation}
To find a particular solution describing the effect of relaxation volumes of defects given by equation \eqref{eq_omega_cylinder}, we note that the divergence of the displacement field generated by $\omega({\bm \rho})$ is given by equation~\eqref{eq_trace_epsilon}, namely
\begin{equation}\label{eq_divergence_displ}
    {\partial \over \partial x_i} u_i({\bm \rho}) = \epsilon_{ii}^{(tot)}({\bf x}) =
            {1\over 3} \left({1+\nu \over 1-\nu}\right) \omega ({\bm \rho}).
\end{equation}
Using the divergence theorem, and noting that the field of displacements depends only on the radial variable, equation \eqref{eq_divergence_displ} gives the particular solution in the form
\begin{equation}\label{particular_cylindrical}
    u_\rho(\rho) = \frac{1}{3\rho}\left(\frac{1+\nu}{1-\nu}\right)
                    \int\limits_{\rho_1}^\rho R\omega(R) \dint{R}.
\end{equation}
The general solution, valid for $\rho_1<\rho<\rho_2$, is given by the sum of the homogeneous solution \eqref{homogeneous_cylindrical} and the particular solution \eqref{particular_cylindrical}, namely   
\begin{equation}\label{eq_displ_cyl}
    u_\rho(\rho)=a\rho+\frac{b}{\rho}+\frac{1}{3\rho}\left(\frac{1+\nu}{1-\nu}\right)\int\limits_{\rho_1}^\rho R\omega
    (R) \dint{R}.
\end{equation}
Combining equations \eqref{elastic_strain}, \eqref{elastic_stress}, and \eqref{eigenstress}, the components of the stress tensor can now be expressed as
\begin{equation*} 
\begin{aligned}
\sigma_{\rho\rho}(\rho) &= 2\mu\epsilon_{\rho\rho}(\rho) + 
                                \frac{2\mu\nu}{1-2\nu}\epsilon_{ii}(\rho) - B\omega(\rho) \\
\sigma_{\varphi\varphi}(\rho) &= 2\mu\epsilon_{\varphi\varphi}(\rho) +  
                                \frac{2\mu\nu}{1-2\nu}\epsilon_{ii}(\rho) - B\omega(\rho) \\
\sigma_{zz}(\rho) &= 2\mu\epsilon_{zz} + 
                                \frac{2\mu\nu}{1-2\nu}\epsilon_{ii}(\rho) - B\omega(\rho),
\end{aligned}
\end{equation*}
%
%
where ${\epsilon_{ii}=\epsilon_{\rho\rho}+\epsilon_{\varphi\varphi}+\epsilon_{zz}}$ is the trace of the elastic strain tensor and $\epsilon_{zz}$ is a constant independent of $\rho$. 

In the unconstrained case, we still have to determine the three constants $a$, $b$, and $c$ in expression \eqref{eq:displ_cyl_general}, derived from equation \eqref{eq_displ_cyl}. Two of the three conditions required for determining the three constants stem from the boundary conditions ${\sigma_{\rho\rho}(\rho_1)=-P}$, where $P$ is the pressure applied to the internal surface, and ${\sigma_{\rho\rho}(\rho_2)=0}$. The third condition can be obtained from the condition of mechanical equilibrium. If a body has an external surface $S$ loaded by external traction forces $t_i$ and is arbitrarily sectioned, then the cut defines an internal section $\Sigma$ with normal vector $n_j$ that divides the external surface into $S_1$ and $S_2$. The equilibrium condition applied to the part of the volume enclosed by $S_1$ and $\Sigma$ implies that 
\begin{equation*}
  \int_{S_1}t_i \text{d}S_1=\int_\Sigma\sigma_{ij}n_j\text{d}\Sigma.  
\end{equation*}
Noting that $t_i=0$ since the pipe is free of external loads, and taking a section perpendicular to $z$, we see that 
\begin{equation*}
    2\pi\int_{\rho_1}^{\rho_2} \sigma_{zz}(R)\text{d}R=0,
\end{equation*}
that is, $\sigma_{zz}$ must be self-equilibrated. Combining the three boundary conditions, we find the radial and axial displacements 
%
%
%
%
%
\begin{align}\label{eq:cyldisplacement_free}
    u_{\rho}(\rho) &=  \frac{P}{2\mu \rho} \frac{\pi\rho_1^2}{A}                             
                            \left[ \rho_2^2 + \left(\frac{1-\nu}{1+\nu}\right) \rho^2 \right] + 
                            \frac{1}{6\rho} \left(\frac{1+\nu}{1-\nu}\right) \nonumber\\
    &\phantom{=} \times \Bigg[
            \rho_1^2 \wmean + \left(\frac{1-3\nu}{1+\nu}\right) \rho^2 \wmean + 2\int_{\rho_1}^{\rho} R\omega(R) \dint{R} \Bigg] \nonumber\\
    u_z(z) &= \left[ 
            \frac{\wmean}{3} - \frac{P}{\mu} \frac{\pi \rho_1^2}{A} \left(\frac{\nu}{1+\nu}\right)
        \right] z
\end{align}
%
where $\wmean$ is the mean value of function $\omega(r)$ over the cross-section area of the cylinder $A = \pi\left(\rho_2^2-\rho_1^2\right)$,
\begin{equation}
  \wmean = \frac{1}{A} \int_S \omega({\bf x}) \dint{S} =\frac{2\pi}{A}\int_{\rho_1}^{\rho_2} R \omega(R)\dint{R}.
\end{equation}

Noting that ${\epsilon_{\rho\rho}=\partial u_\rho/\partial \rho}$, ${\epsilon_{\varphi\varphi}=u_\rho/\rho}$, and ${\epsilon_{zz}=\partial u_z/\partial z}$, we find the three non-vanishing components of stress 
%
%
%
%
%
%
%
\begin{equation}
\begin{aligned}
    &\sigma_{\rho\rho}(\rho) = \frac{\pi\rho_1^2 P}{A} \left(1-\frac{\rho_2^2}{\rho^2}\right)
                        +\frac{\mu}{3} \left(\frac{1+\nu}{1-\nu}\right)\times  \\
    &\phantom{=}\left[
            \left(1-\frac{\rho_1^2}{\rho^2}\right) \wmean - \frac{2}{\rho^2}\int_{\rho_1}^{\rho} R\omega(R)\text{d}R 
    \right]
    \label{eq_Srr_cyl}
\end{aligned}
\end{equation}
\begin{equation}\label{eq_Stt_cyl}
\begin{aligned}
    &\sigma_{\varphi\varphi}(\rho) = \frac{\pi\rho_1^2 P}{A} \left(1+\frac{\rho_2^2}{\rho^2}\right)
                                        +\frac{\mu}{3} \left(\frac{1+\nu}{1-\nu}\right)\times \\
    &\phantom{=}\left[
            \left(1+\frac{\rho_1^2}{\rho^2}\right) \wmean + \frac{2}{\rho^2}\int_{\rho_1}^{\rho} R\omega(R)\text{d}R - 2\omega(\rho)
      \right]
\end{aligned}
\end{equation}
\begin{equation}\label{eq:Szz_cyl}
    \sigma_{zz}(\rho) = \frac{2\mu}{3} \left(\frac{1+\nu}{1-\nu}\right) 
            \big[\wmean - \omega(\rho)\big].
\end{equation}

The above equations show that $\sigma_{\rho\rho}(\rho)$ in \eqref{eq_Srr_cyl} vanishes at $\rho =\rho_1$ and $\rho=\rho_2$ for an arbitrary distribution $\omega(\rho)$. Because of the linear superposition principle, effects of internal pressure and irradiation are completely decoupled. In the absence of irradiation, the expressions for the components of stress readily reduce to the Lam\'e equations for a thick-walled pressurised cylinder \cite{Perry2003}. If swelling, although present, is spatially homogeneous, its contribution to all the stress components vanishes identically, as expected for an unconstrained body subject to traction-free boundary conditions.

If the tube is constrained along $z$, so that ${u_z=0}$, ${\epsilon_{zz}=0}$, then only two conditions need to be specified to determine the constants in equation \eqref{eq_displ_cyl}. It is sufficient to prescribe the values of the radial stress at internal and external surfaces, namely ${\sigma_{\rho\rho}(\rho_1)=-P}$ and ${\sigma_{\rho\rho}(\rho_2)=0}$. In fact, the presence of external forces, required to keep the length of the tube constant, means that $\sigma_{zz}$ need not to self-equilibrate. The radial displacement function in the constrained ($C$) case is 
%
%
%
%
%
\begin{align}\label{eq:cyldisplacement}
    u_{\rho}^C(\rho) &= \frac{P}{2\mu \rho} \frac{\pi\rho_1^2}{A} \left[\rho_2^2 + (1-2\nu)\rho^2\right] 
        + \frac{1}{6\rho} \left(\frac{1+\nu}{1-\nu}\right)  \nonumber\\
    &\times \left[
            \rho_1^2 \wmean + (1-2\nu)\rho^2 \wmean 
                + 2\int_{R_1}^{\rho} R\omega(R)\dint{R} 
        \right].
\end{align}
We find that the stresses $\sigma_{\rho\rho}(\rho)$ and $\sigma_{\varphi\varphi}(\rho )$ are not affected by the different boundary conditions and remain the same as in equations \eqref{eq_Srr_cyl} and \eqref{eq_Stt_cyl}. $\sigma_{zz}(\rho )$ in the constrained case, on the other hand, has the form
%
%
%
%
\begin{equation}\label{eq:Szz_cyl_C}
    \sigma_{zz}^C(\rho) = \frac{2 \pi \nu \rho_1^2  P}{A} 
                        + \frac{2\mu}{3} \left(\frac{1+\nu}{1-\nu}\right) 
                            \big[ \nu\wmean -\omega(\rho) \big].
\end{equation}
A comparison of equations \eqref{eq:Szz_cyl} and \eqref{eq:Szz_cyl_C} highlights an important point. Spatially homogeneous irradiation generates no $\sigma_{zz}$ component of elastic stress if the pipe is unconstrained, but if the extremities are fixed and $P=0$, then we have
\begin{equation}
    \sigma_{zz}^C = -\frac{2\mu}{3} (1+\nu) \overline{\omega}.
\end{equation}
The above equations shows that the magnitude of elastic stress generated by irradiation is of the order of $\omega$ times the shear modulus. As the former can be as large as \SI{1}{\percent} or more, see \cite{Garner2020,Zinkle2014}, internal elastic stress reaching hundreds of MPa can develop in the absence of any internal gas pressure in a pipe with constrained ends, even in the limit where irradiation is spatially homogeneous.

\begin{figure*}[p]
  \includegraphics[width=0.4\textwidth]{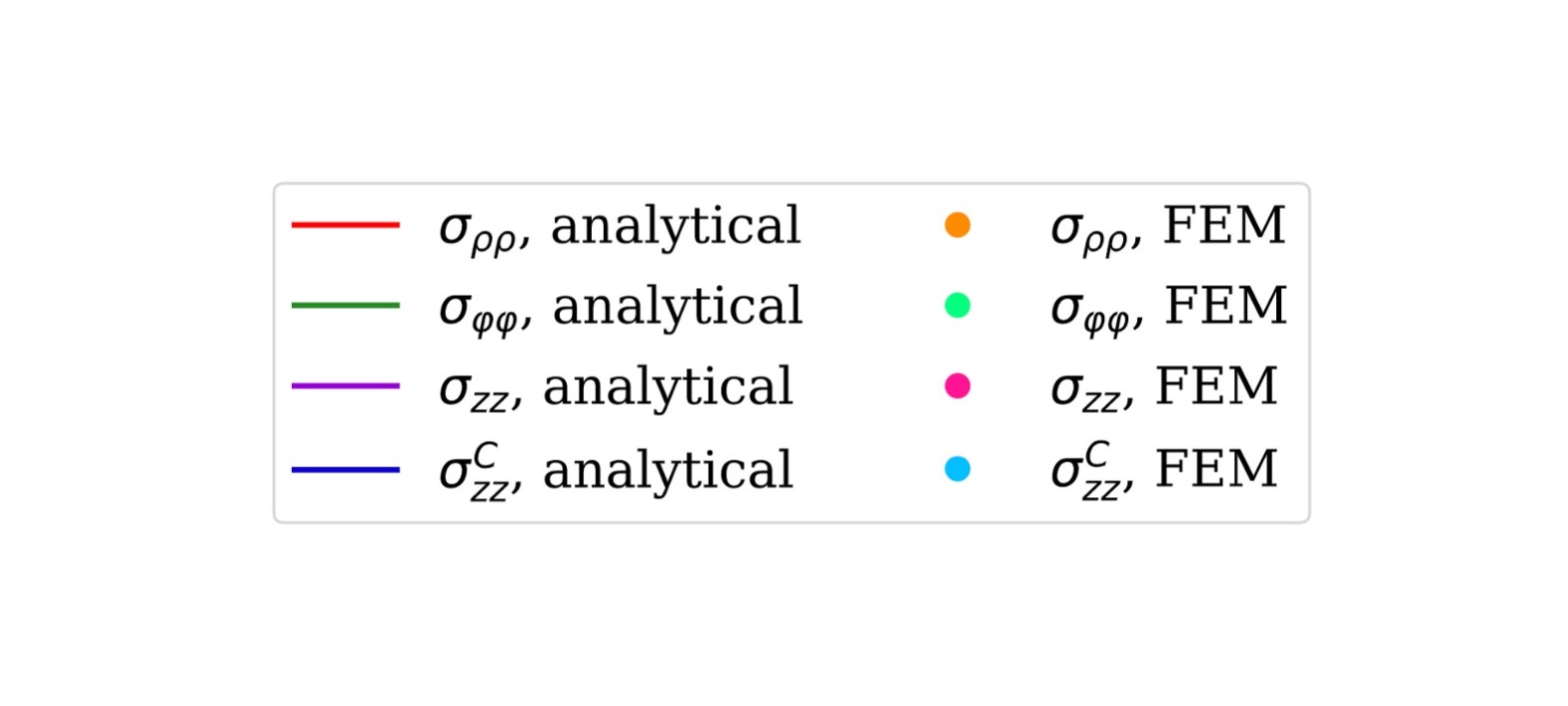}%
\subfloat[\label{fig:omega_cyl}]{%
  \includegraphics[width=0.3\textwidth]{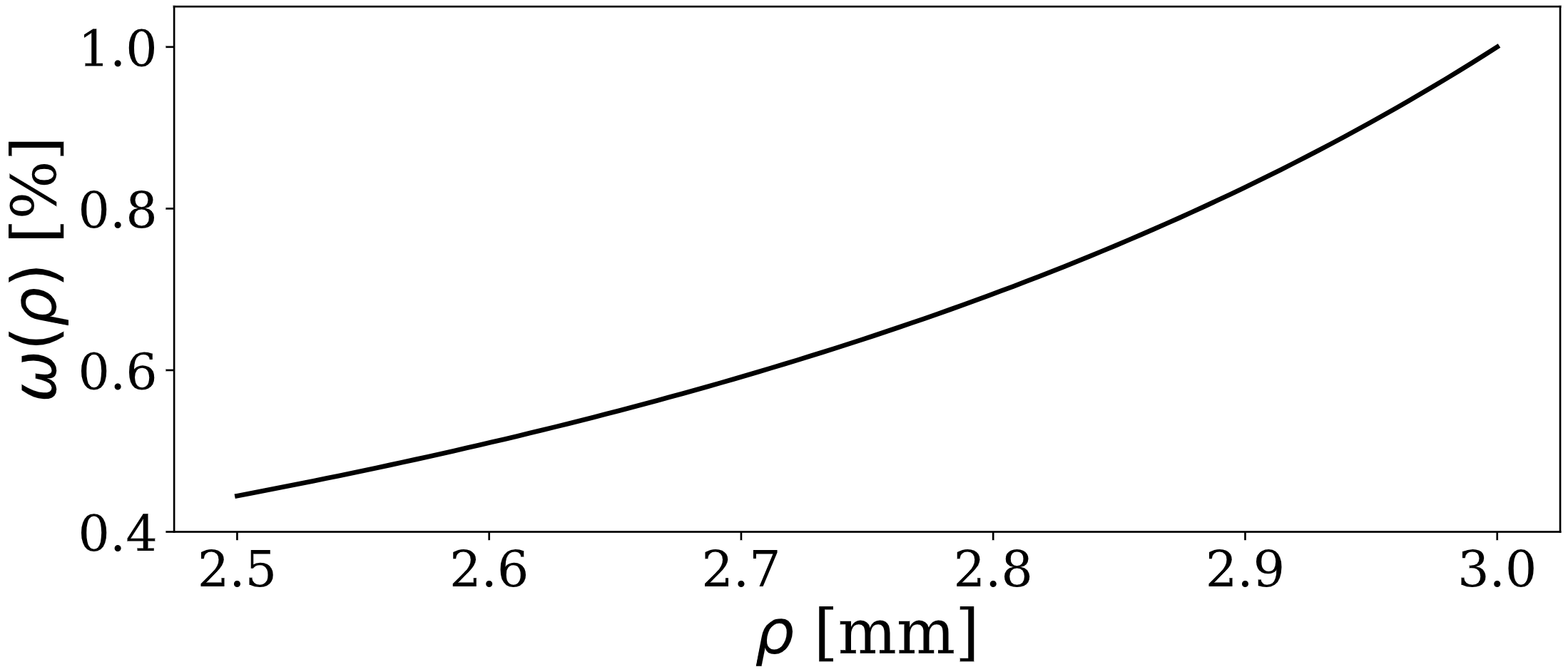}%
}\hfill
\subfloat[\label{fig:str_p}]{%
  \includegraphics[width=0.45\textwidth]{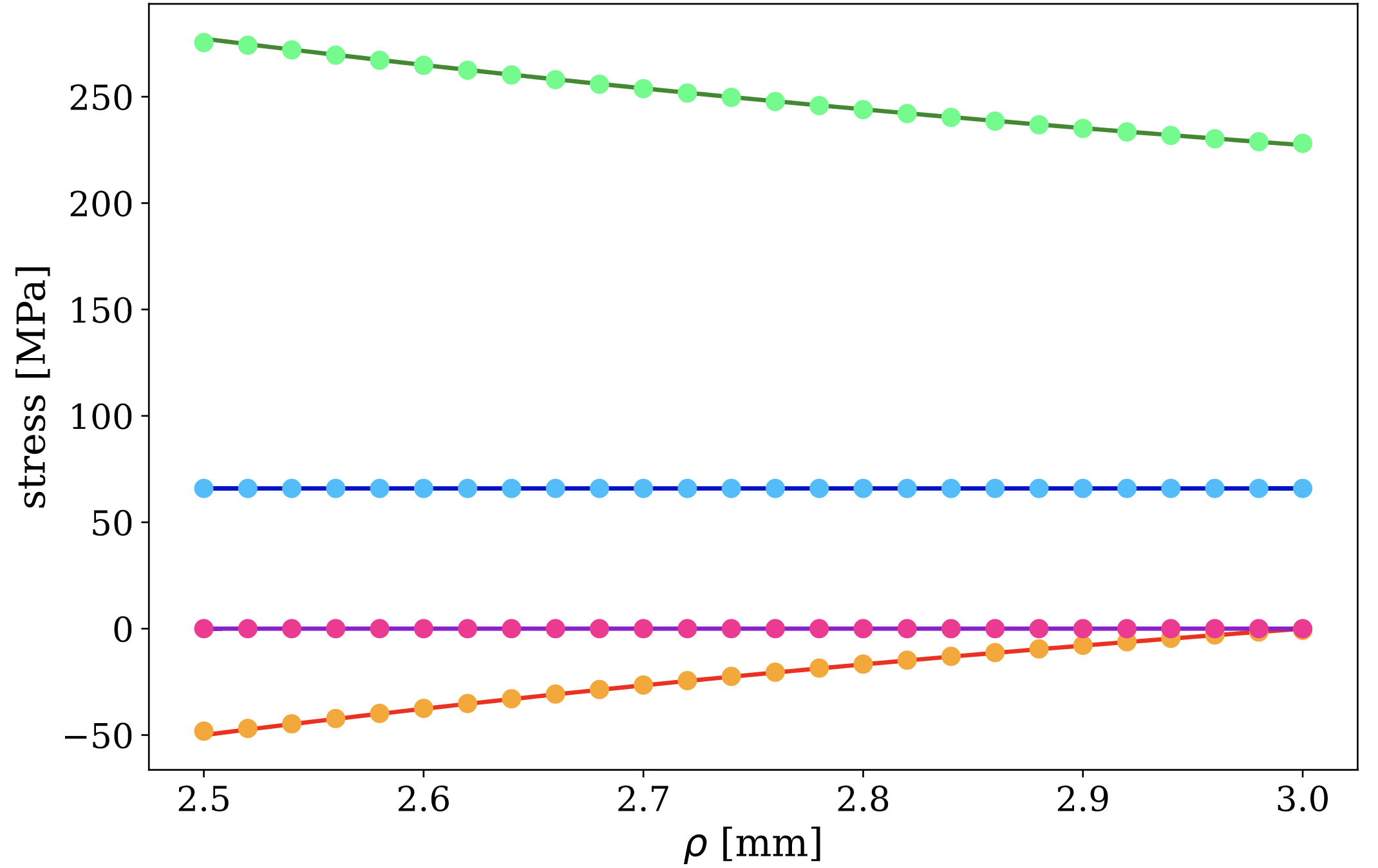}%
}\hfill
\subfloat[\label{fig:FEM_p}]{%
  \includegraphics[width=0.4\textwidth]{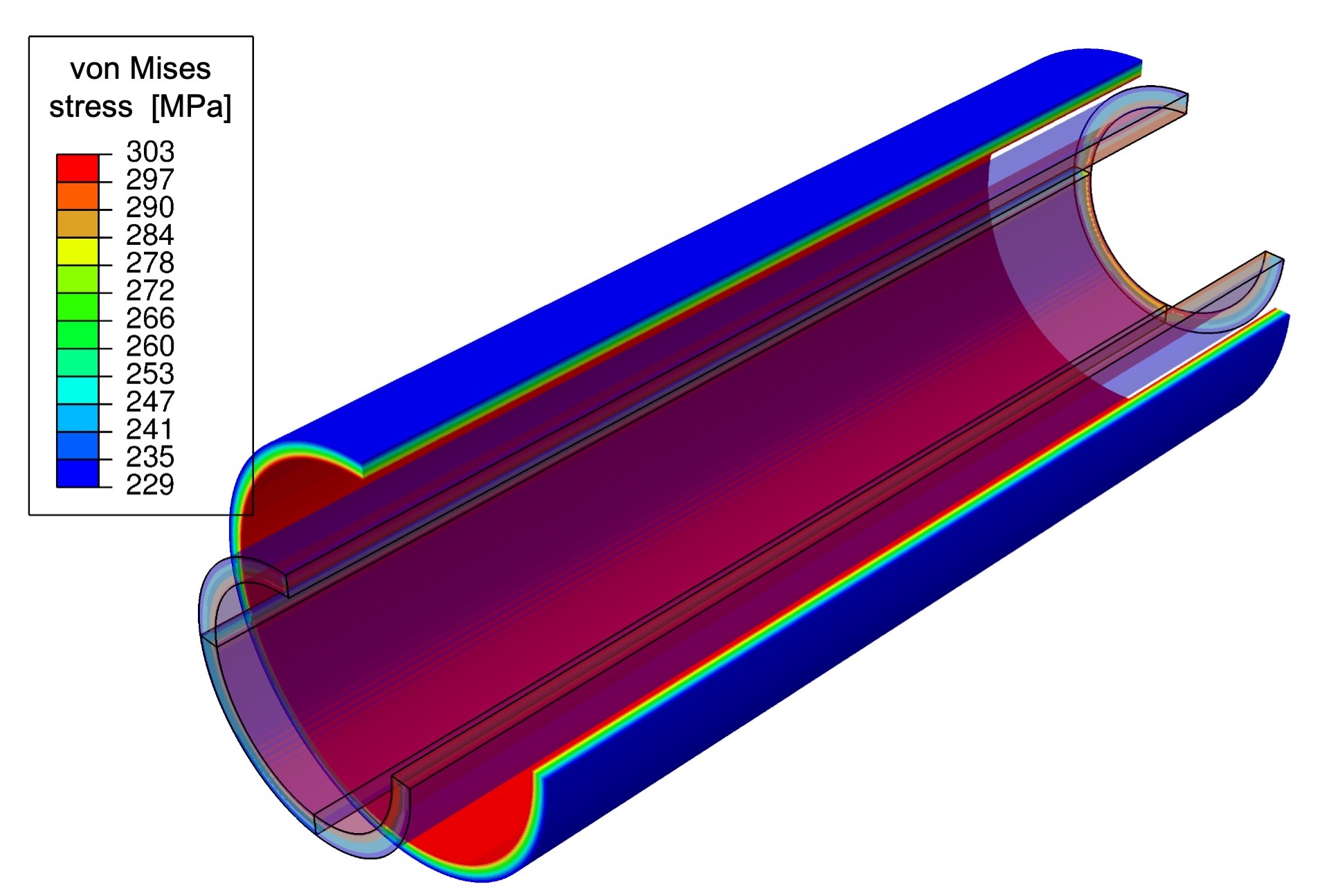}%
}\hfill
\subfloat[\label{fig:str_sw}]{%
  \includegraphics[width=0.45\textwidth]{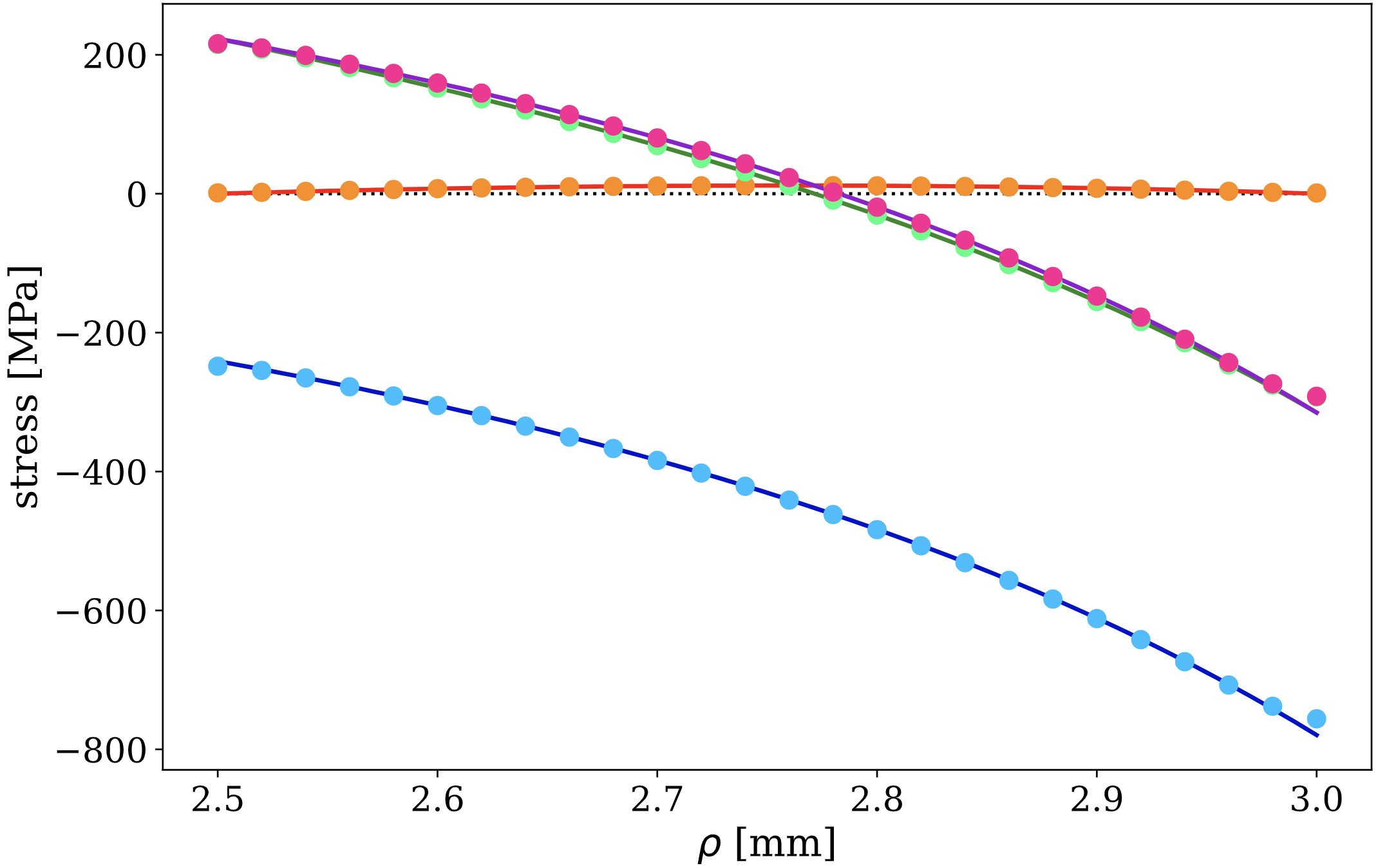}%
}\hfill
\subfloat[\label{fig:FEM_sw}]{%
  \includegraphics[width=0.4\textwidth]{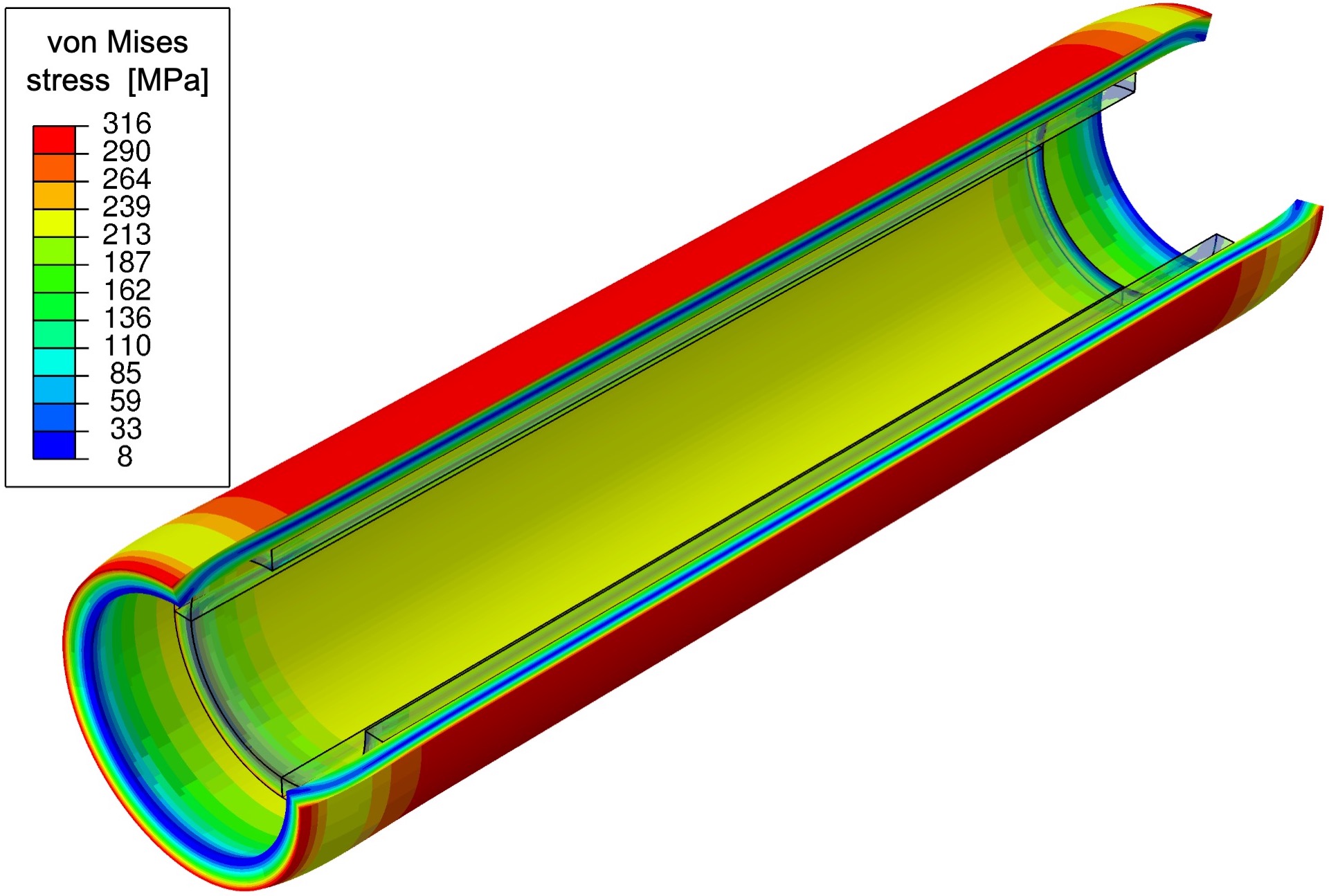}%
}\hfill
\subfloat[\label{fig:str_p_sw}]{%
  \includegraphics[width=0.45\textwidth]{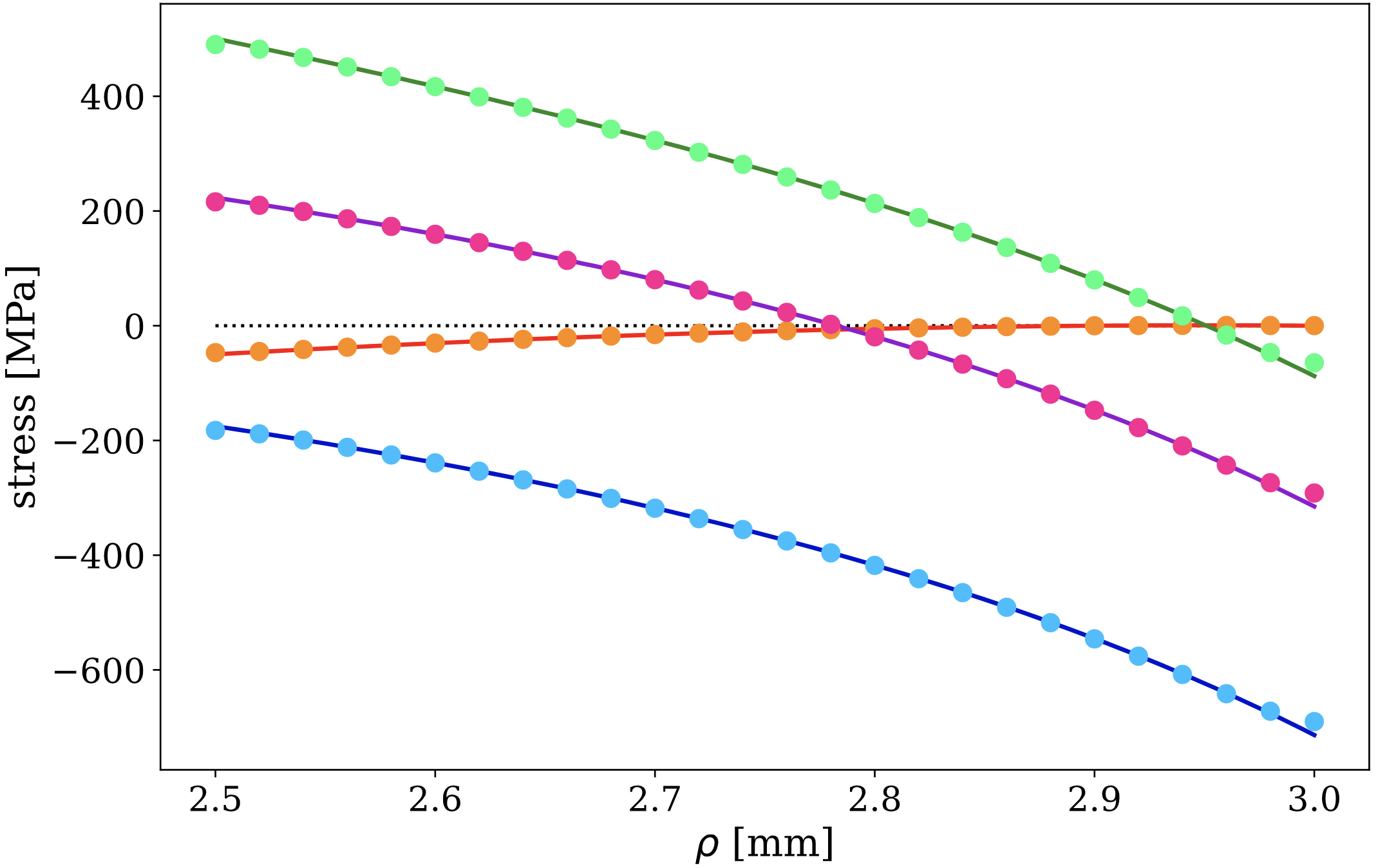}%
}\hfill
\subfloat[\label{fig:FEM_p_sw}]{%
  \includegraphics[width=0.4\textwidth]{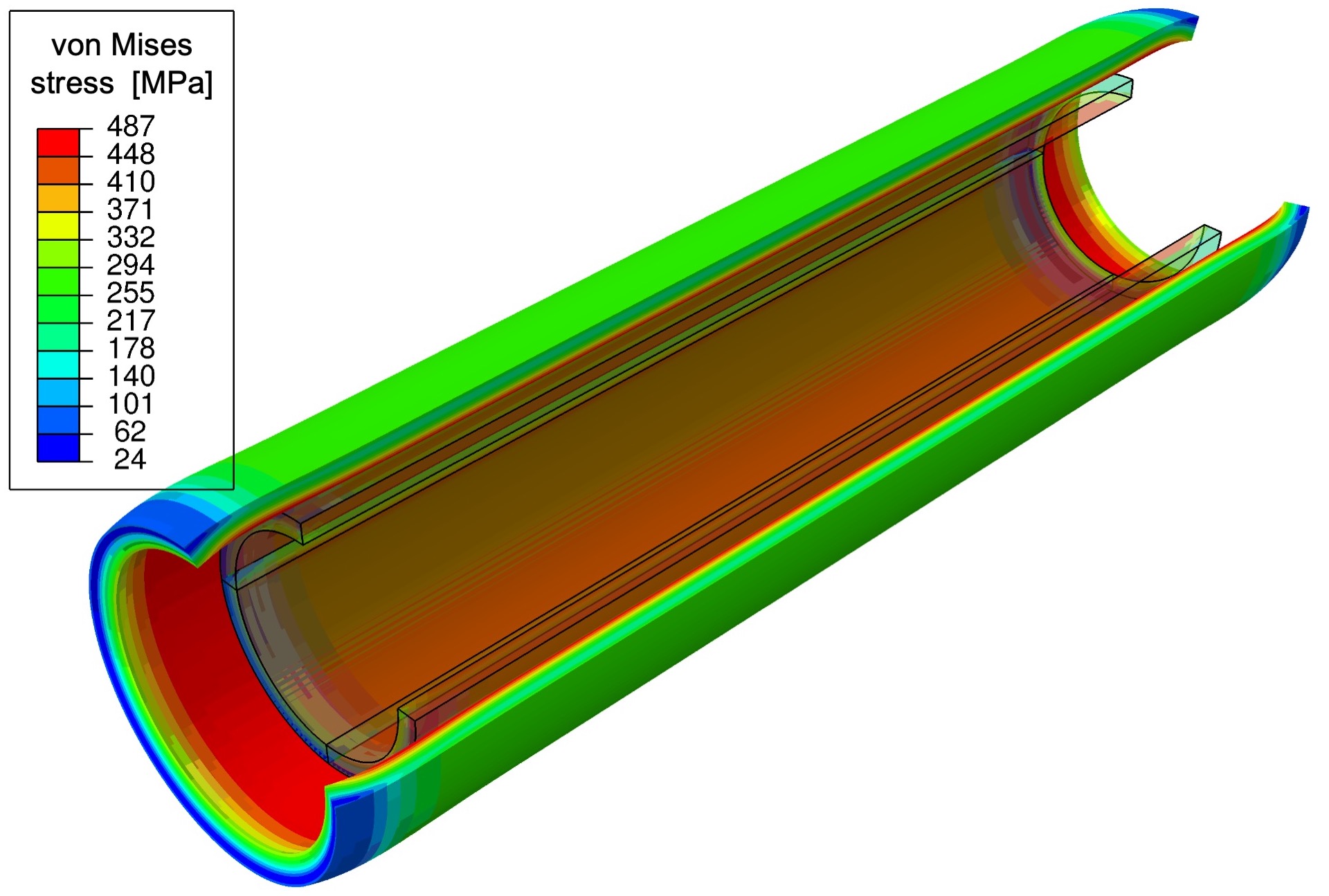}%
}\hfill
    \caption{Summary of the stress analysis for a pressurised and irradiated tube. In (b) and (c) only pressure is present, in (d) and (e) only irradiation, in (f) and (g) both are acting. Next to the corresponding plot, the von Mises stress is shown on a sectioned view of the component for the unconstrained case (the deformations are exaggerated for clarity and the undeformed shape is shaded). The assumed profile of $\omega(\rho)$ is shown in (a), the inner pressure is 50 MPa. The analytical solution (solid lines) is in agreement with the FEM model (dots).}
    \label{fig:cyl}
\end{figure*}

Dimensional stability is an important consideration in the context of design and operation of a nuclear reactor involving many interdependent components. As a result of applied pressure and irradiation, the volume of a pipe can change. This volume change is given by the volume integral of the trace of the total strain. If the pipe is free then, far from its extremities, the volume change of the pipe per unit length can be computed as
%
%
%
\begin{align}
   \frac{\Delta V}{L}   &= \frac{1}{L} \int_V \frac{\partial u_i}{\partial x_i}\, \dint{V}    
                         = 2\pi\int_{\rho_1}^{\rho_2} \left(
                            \frac{\partial (\rho u_\rho)}{\partial \rho}
                            +\rho\frac{\partial u_z}{\partial z}\right) \dint{\rho} \nonumber\\
                        &= \frac{P}{\mu} \left(\frac{1-2\nu}{1+\nu}\right) \pi\rho_1^2         
                            + A \wmean.
   \label{eq:deltaV_cyl}
\end{align}
If the pipe is constrained, the change of volume is given by
%
%
\begin{align}
  \frac{\Delta V^C}{L}  &= \frac{1}{L} \int_V \frac{\partial u^C_i}{\partial x_i} \dint{V} \nonumber \\
                        &= \frac{P}{\mu}(1-2\nu) \pi\rho_1^2 
                        + \frac{2}{3}(1+\nu)A \wmean.
  \label{eq:deltaV_cyl_C}
\end{align}
Equation \eqref{eq:deltaV_cyl} shows that in the limit where $P=0$, the volume change of an unconstrained irradiated pipe equals the total relaxation volume of all the defects in the material. If the pipe is constrained, the volume change is smaller than the total relaxation volume of defects, and the difference can be attributed to the effect of external constraints. Both equations \eqref{eq:deltaV_cyl} and \eqref{eq:deltaV_cyl_C} confirm the intuitive conclusion that in the limit where the material is  incompressible (${\nu=1/2}$), the application of external pressure $P$ does not alter the volume of the pipe.

Equations \eqref{eq:deltaV_cyl} and \eqref{eq:deltaV_cyl_C} can also be derived by evaluating, to first order, the change of volume associated with the transformation of a cylinder from its initial undeformed configuration to the final one with the inner radius of ${\rho_1+u_\rho(\rho_1)}$, outer radius of ${\rho_2+u_\rho(\rho_2)}$ and the length of ${L+u_z(L)}$.


An FEM model was constructed for a pipe made of ferritic steel ($\mu=\SI{80}{\GPa}$, $\nu=0.29$), with $\rho_1=\SI{2.5}{mm}$, $\rho_1=\SI{3.0}{mm}$ and pressure of $P=\SI{50}{\MPa}$ acting on its inner surface. These parameters are close to the experimental conditions for pressurised steel specimens used in irradiation tests \cite{Pintsuk2019}. Function $\omega(\rho)$ was selected in such a way so that the density of relaxation volumes varies from about 1\% at the outer surface, directly exposed to irradiation, to about 0.5\% at the inner surface, see equation \eqref{eq:om_cyl_FEM}. Numerical FEM solutions are illustrated in Fig. \ref{fig:cyl}. Simulations were performed assuming the effect of pressure only, irradiation only, and the combined effect of both pressure and irradiation. To achieve agreement between numerical FEM results and exact analytical solutions, about $9.40\times10^5$ hexahedral linear elements were used.

If the exposure of a pipe to an external source of irradiation gives rise to a spatially varying defect density, the maximum tensile hoop stress, $\sigma_{\varphi\varphi}$, occurs at the inner surface of the pipe, which is also where the hoop stress induced by $P$ is maximum. The axial stress, $\sigma_{zz}$, is negative throughout the thickness of the pipe if it is constrained. If the pipe can expand freely along its length, then the axial stress must change sign to satisfy the condition of self-equilibrium. A change in the density of relaxation volumes of 0.5\% produces stresses that are comparable with those generated by high internal pressure, with potential implication for the design of experimental tests.


\section{elastic Stress and strain in a spherical shell exposed to irradiation}\label{sec:sphere}
In this section we evaluate elastic stress and strain developing in a spherical shell exposed to irradiation. The advantage offered by the spherical symmetry of the shell is that in the limit where the source of irradiation is also spherically symmetric, the solutions can be found in a closed analytical form. Some of them, related to the total strain and stress, were investigated earlier in Ref. \cite{NF2018}, however the pure elastic components of stress and strain were not evaluated. Given the significance of pure elastic solutions in the context of assessment of structural integrity of power plant components, we present the relevant analysis below.  

If the source of irradiation is spherically symmetric, the distribution of defect relaxation volumes depends only on the distance $r$ to the centre of the shell. It is independent of the polar and azimuthal angles $\theta$ and $\phi$ of the spherical system of coordinates, the origin of which is at the centre of the shell. The density of relaxation volume tensors now has the form
\begin{equation}
    \omega_{mn}({\bf x}) = {1\over 3} \omega(r) \delta_{mn},
\end{equation}
and the field of displacements is defined on the interval $R_1\le r \le R_2$, where $R_1$ is the inner radius of the shell and $R_2$ is its outer radius. This field is radially-symmetric and the vector of displacements can be written as ${\bf u}({\bf r})=u_r(r){\bf n}$, where ${\bf n}={\bf r}/r$. The divergence of the field of atomic displacements has the form \eqref{eq_trace_epsilon}
\begin{equation}
    {\partial \over \partial r} u(r)= {1\over 3}\left({1+\nu \over 1-\nu}\right)\omega (r).
    \label{displacements_divergence}
\end{equation} 
Since the field of displacements is radially symmetric, we apply the divergence theorem to equation \eqref{displacements_divergence} and write
\begin{equation}\label{u_r}
4\pi r^2 u_r(r) = {4\pi\over 3} \left({1+\nu \over 1-\nu}\right) \int_{R_1}^r R^2\omega(R) \dint{R},
\end{equation}
for $R_1 \le r \le R_2$. 

Noting that in the absence of radiation defects the divergence of ${\bf u}({\bf x})$ is a harmonic function of coordinates, we write the field of displacements as a sum of the particular solution of heterogeneous equation \eqref{displacements_divergence} and a general solution of the corresponding homogeneous equation, namely
\begin{equation}\label{u_r_full}
    u_r(r) = ar + {b\over r^2} + {1\over 3r^2}\left({1+\nu \over 1-\nu}\right) 
                \int_{R_1}^r R^2 \omega(R) \dint{R}.
\end{equation}
Here $a$ and $b$ are constants that need to be determined from the boundary conditions at $r=R_1$ and $r=R_2$. Assuming that pressure at $R_1$ and $R_2$ is negligible in comparison with the stresses developing in the material due to the accumulation of defects, we adopt the traction-free boundary conditions $\sigma_{ij}(r)n_j=0$ at $r=R_1$ and $r=R_2$.
The total strain can be found by differentiating \eqref{u_r_full}. It has the form
\begin{align}
    \epsilon_{ij}^{(tot)}(r) &= a\delta_{ij} + {b\over r^3} \left(\delta _{ij} - 3 n_i n_j\right)\nonumber \\
    &+ {1\over 3r^3}\left({1+\nu \over 1-\nu}\right)
       \left(\delta_{ij} - 3 n_i n_j\right) \int_{R_1}^r R^2 \omega(R) \dint{R}\nonumber \\
    &+ {1\over 3} \left({1+\nu \over 1-\nu}\right) n_i n_j \omega(r).
    \label{strain_full}
\end{align}
The find the pure elastic part of strain, we need to subtract the eigenstrain of defects ${1\over 3}\omega(r) \delta_{ij}$ from the above expression, resulting in  
\begin{align}
    \epsilon_{ij}(r) &= a\delta_{ij} + {b\over r^3} \left(\delta_{ij} - 3 n_i n_j\right)\nonumber \\
    &+ {1\over 3r^3} \left({1+\nu \over 1-\nu}\right)
       \left(\delta_{ij} - 3 n_i n_j\right) \int_{R_1}^r R^2 \omega(R)\dint{R}\nonumber \\
    &+ {1\over 3}\left[\left({1+\nu \over 1-\nu}\right) n_i n_j -\delta_{ij}\right] \omega(r).
    \label{strain_elastic}
\end{align}
To find the pure elastic stress, we multiply $\epsilon_{ij}(r)$ by the forth-rank tensor of elastic constants \eqref{elastic_constants}. The resulting expression for elastic stress is
\begin{align}
    \sigma_{ij}(r) &= 2\mu a\left({1+\nu \over 1-2\nu}\right)\delta_{ij} + 
                        {2\mu b\over r^3}(\delta_{ij}-3n_in_j)\nonumber \\
    &+ {2\mu \over 3r^3}\left({1+\nu \over 1-\nu}\right)
            \left(\delta _{ij}-3n_in_j\right)\int_{R_1}^r R^2 \omega(R) \dint{R}\nonumber \\
    &+ {\mu\over 3} \left({1+\nu \over 1-\nu}\right) \omega (r) \left( {2\nu \over 1-2\nu} \delta_{ij}
            + 2 n_i n_j\right)\nonumber \\
    &- {2\mu \over 3}\left({1+\nu \over 1-2\nu} \right)\omega(r)\delta_{ij}.
    \label{stress_elastic}
\end{align}
Projecting elastic stress onto the radial unit vector ${\bf n}={\bf r}/r$, we find
\begin{align}
    \sigma_{ij}(r)n_j &= 2\mu a \left({1+\nu \over 1-2\nu}\right)n_i - {4\mu b\over r^3}n_i\nonumber \\
                      & -{4\mu \over 3r^3}\left({1+\nu \over 1-\nu}\right) n_i \int_{R_1}^r R^2 \omega(R) \dint{R}.
    \label{stress_elastic_projection}
\end{align}
Applying the traction-free boundary conditions $\sigma_{ij}(r)n_j=0$ at $r=R_1$ and $r=R_2$ to \eqref{stress_elastic} and \eqref{stress_elastic_projection}, we find parameters $a$ and $b$, namely 
%
%
%
%
%
\begin{equation}\label{a}
    a = \frac{2}{9} \left(\frac{1-2\nu}{1-\nu}\right) \wmean
\end{equation}
and
\begin{equation}\label{b}
    b = \frac{1}{9} \left(\frac{1+\nu}{1-\nu}\right) \wmean,
\end{equation}
where $\wmean$ is the mean density of relaxation volumes of all the defects accumulated in the shell
\begin{equation}
  \wmean = \frac{1}{V} \int_V \omega({\bf x}) \dint{V} =\frac{4\pi}{V}\int_{R_1}^{R_2} R^2 \omega(R)\dint{R},
\end{equation}
and $V$ is the geometric volume of the shell
\begin{equation}
    V = \frac{4\pi}{3} \left(R_2^3 - R_1^3\right).
\end{equation}
%
%
The above formulae remain valid irrespectively of whether or not $\omega (r)$ vanishes at surfaces $r=R_1$ and $r=R_2$.

The total macroscopic change of volume resulting from the accumulation of defects in the the shell equals the integral of the trace of the {\it total} strain tensor \eqref{total_volume} over the volume of the component
\begin{align}
    \Delta V    &= \int_V \epsilon_{ii}({\bf r})\dint{V} 
                 = \int_V \left[{1\over 3}\left({1+\nu \over 1-\nu}\right)
                        \omega({\bf r}) + 3a\right] \dint{V}\nonumber \\
                &= \int_{R_1}^{R_2} \left[{1\over 3} \left({1+\nu \over 1-\nu}\right) 
                        \omega(R)+3a\right]4\pi R^2 \dint{R},
    \label{swelling}
\end{align}
where we noted that $\delta_{ii}=3$ and $n_in_i=1$. The integral equals $\Omega_{tot} = V \wmean$, in agreement with equation \eqref{total_volume}. 

\begin{figure}
\subfloat[\label{fig:sph_omega}]{%
  \includegraphics[width=0.45\textwidth]{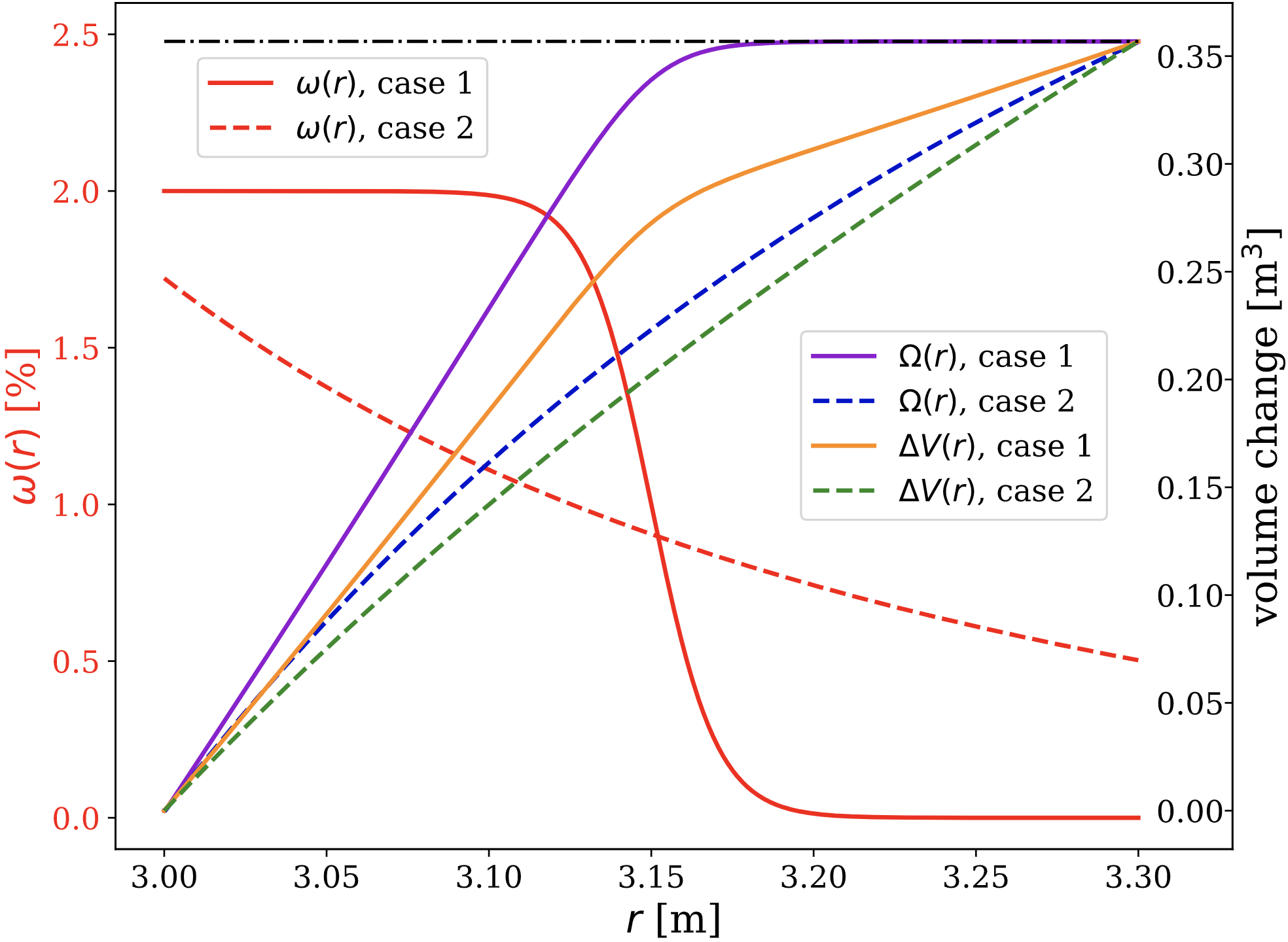}%
}\hfill
\subfloat[\label{fig:sph_stresses}]{%
  \includegraphics[width=0.45\textwidth]{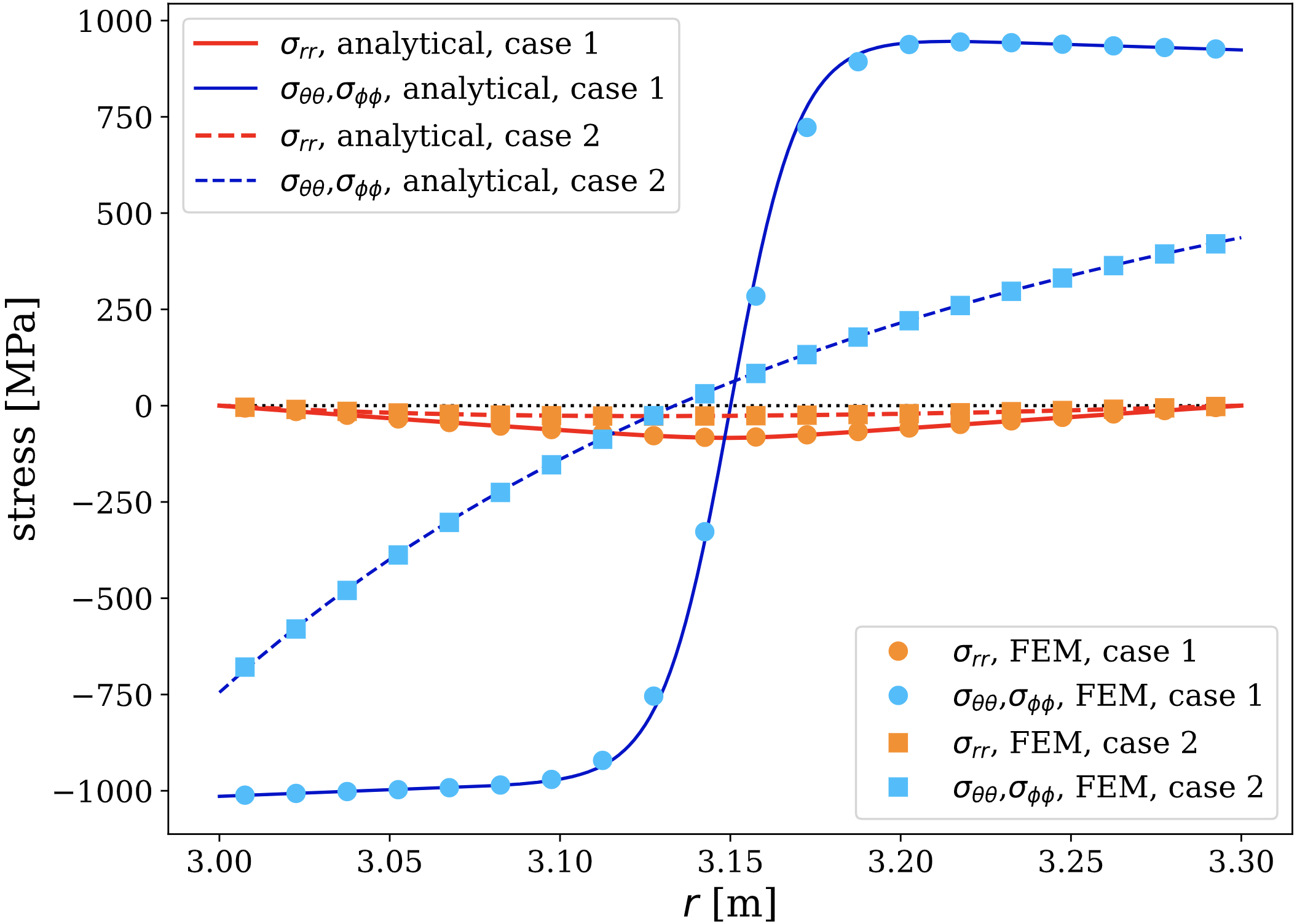}%
}\hfill
\subfloat[\label{fig:sph_displ}]{%
  \includegraphics[width=0.45\textwidth]{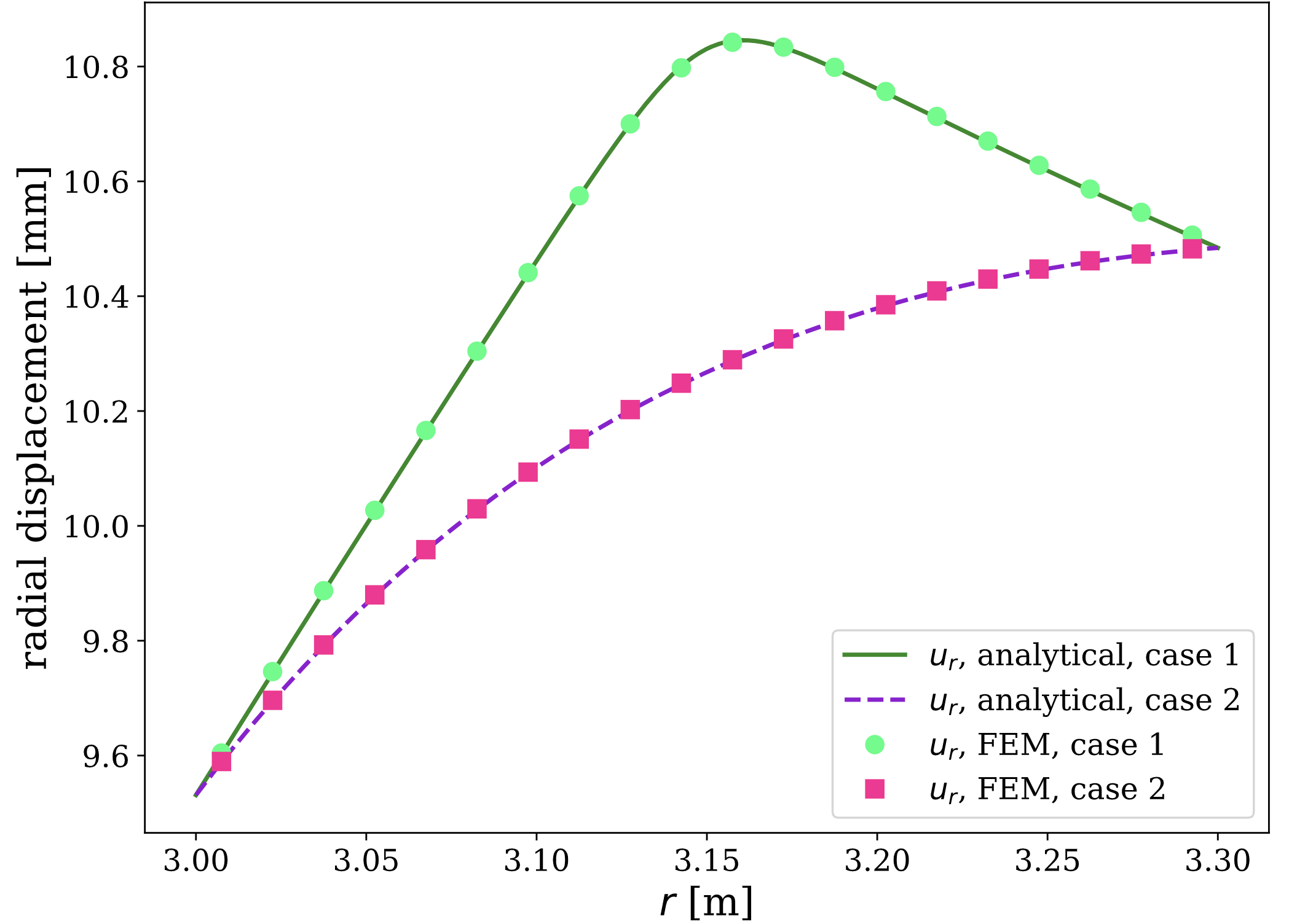}%
}\hfill
    \caption{(a) The two profiles of $\omega(r)$ used in the analysis of solutions for the spherical shell (case 1 and 2). Integral $\Omega(r)=4\pi\int_{R_1}^r\omega(r)R^2\text{d}R$ equals the total volume of all the defects in the shell. Cases 1 and 2 were chosen to have the identical values of $\Omega_{tot}=\Omega(R_2)$. $\Omega_{tot}$ also equals the change of volume found by evaluating the integral of the total strain, $\Delta V(r)=4\pi\int_{R_1}^r\epsilon_{ii}R^2\text{d}R$, at $r=R_2$, compared to the analytical result (\ref{swelling_surfaces}), shown by a black line. (b) Stresses depend on the shape of the distribution $\omega(r)$ and reach fairly high values. (c) On the other hand, the total deformation of the shell is given by the integral $\int\omega ({\bf x}) \text{d}V$, and for the same value of this integral, $u_r(R_1)$ and $u_r(R_2)$ are the same. The analytical and FEM solutions are found to be in full agreement.}
    \label{fig:sphere_plots}
\end{figure}

Substituting \eqref{a} and \eqref{b} into \eqref{u_r_full}, we find the radial displacements of the inner and outer surfaces of the shell 
%
%
%
%
\begin{equation}\label{inner_and_outer_surfaces}
\begin{aligned}
    u_r(R_1) &= {1\over 3} R_1 \wmean \\
    u_r(R_2) &= {1\over 3} R_2 \wmean.
\end{aligned}
\end{equation}
These displacements satisfy the condition
\begin{equation}\label{swelling_surfaces}
    4\pi R^2_2u_r(R_2)-4\pi R_1^2u_r(R_1) = V\wmean = \Omega_{tot},
\end{equation}
which provides an alternative way of evaluating the total volumetric swelling of the shell \cite{NF2018}. Fig. \ref{fig:sph_omega} graphically exemplifies equation \eqref{total_volume} and is in full agreement with \eqref{swelling_surfaces}, for two functions $\omega(r)$ that have the same $\wmean$.

To assess the performance of a component under operating conditions where radiation defects accumulate in the bulk of its structure, it is convenient to use the spherical system of coordinates and project the elastic stress tensor \eqref{stress_elastic} onto the three orthogonal unit vectors $({\bf e}_r, {\bf e}_{\theta},{\bf e}_{\phi})$, corresponding to the radial and angular degrees of freedom and related to the various modes of deformation of the shell. These unit vectors are
\begin{equation} 
\begin{aligned}
    {\bf e}_r &=    {\bf e}_x\sin \theta \cos \phi +
                    {\bf e}_y\sin \theta \sin \phi +
                    {\bf e}_z\cos \theta  \\
    {\bf e}_{\theta} &=     {\bf e}_x\cos \theta \cos \phi +
                            {\bf e}_y\cos \theta \sin \phi -
                            {\bf e}_z\sin \theta \\
    {\bf e}_{\phi}&=   -{\bf e}_x \sin \phi +
                        {\bf e}_y \cos \phi, 
    \label{unit_vectors_spherical}
\end{aligned}
\end{equation}
where $\theta$ and $\phi$ are the polar and azimuthal angles of the spherical system of coordinates. Since ${\bf e}_r$ is the same as ${\bf n}$, we have $({\bf n}\cdot {\bf e}_{\theta})=({\bf n}\cdot {\bf e}_{\phi})=0.$

The radial diagonal element of the elastic stress tensor is
%
%
%
\begin{equation}\label{stress_rr}
\begin{aligned}
    \sigma_{rr}(r)  &= {4\mu \over 9} \left({1+\nu \over 1-\nu}\right)                                      
                            \left[\wmean\left(1-{R_1^3\over r^3}\right)\right]    \\
                    &- {4\mu \over 9}\left({1+\nu \over 1-\nu}\right)
                            \left[{3 \over r^3}\int_{R_1}^r R^2\omega(R) \dint{R}\right].
\end{aligned}
\end{equation}
A direct calculation shows that in the limit where function $\omega(r)$ does not depend on $r$, $\omega(r)=const$, the radial component of elastic stress vanishes identically everywhere in the volume of the shell, $\sigma _{rr}(r)=0$ for all $R_1<r<R_2$.

\begin{figure}[h]
\subfloat[\label{fig:sph_FEM_1}]{%
  \includegraphics[width=0.42\textwidth]{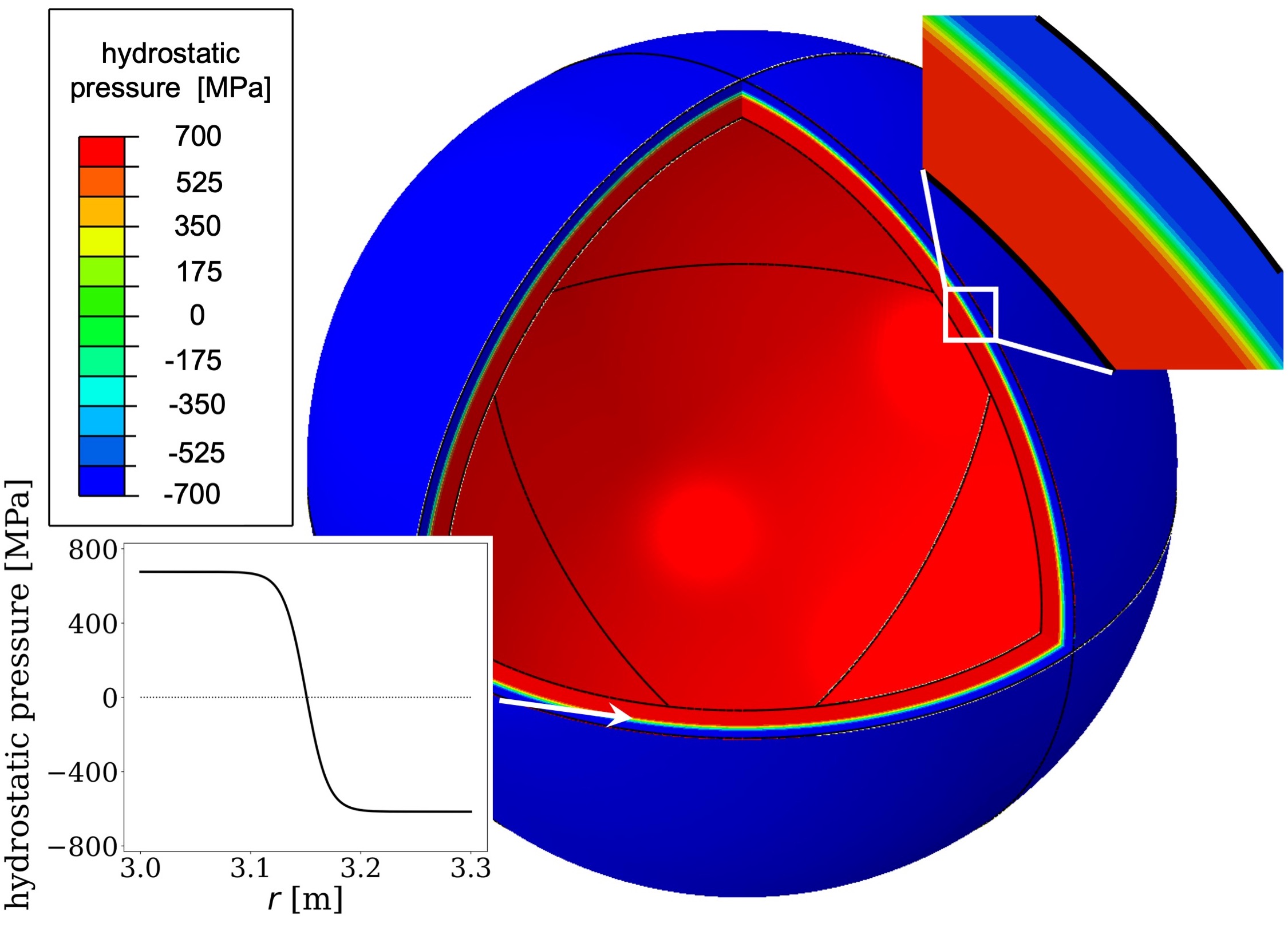}%
}\hfill
\subfloat[\label{fig:sph_FEM_2}]{%
  \includegraphics[width=0.42\textwidth]{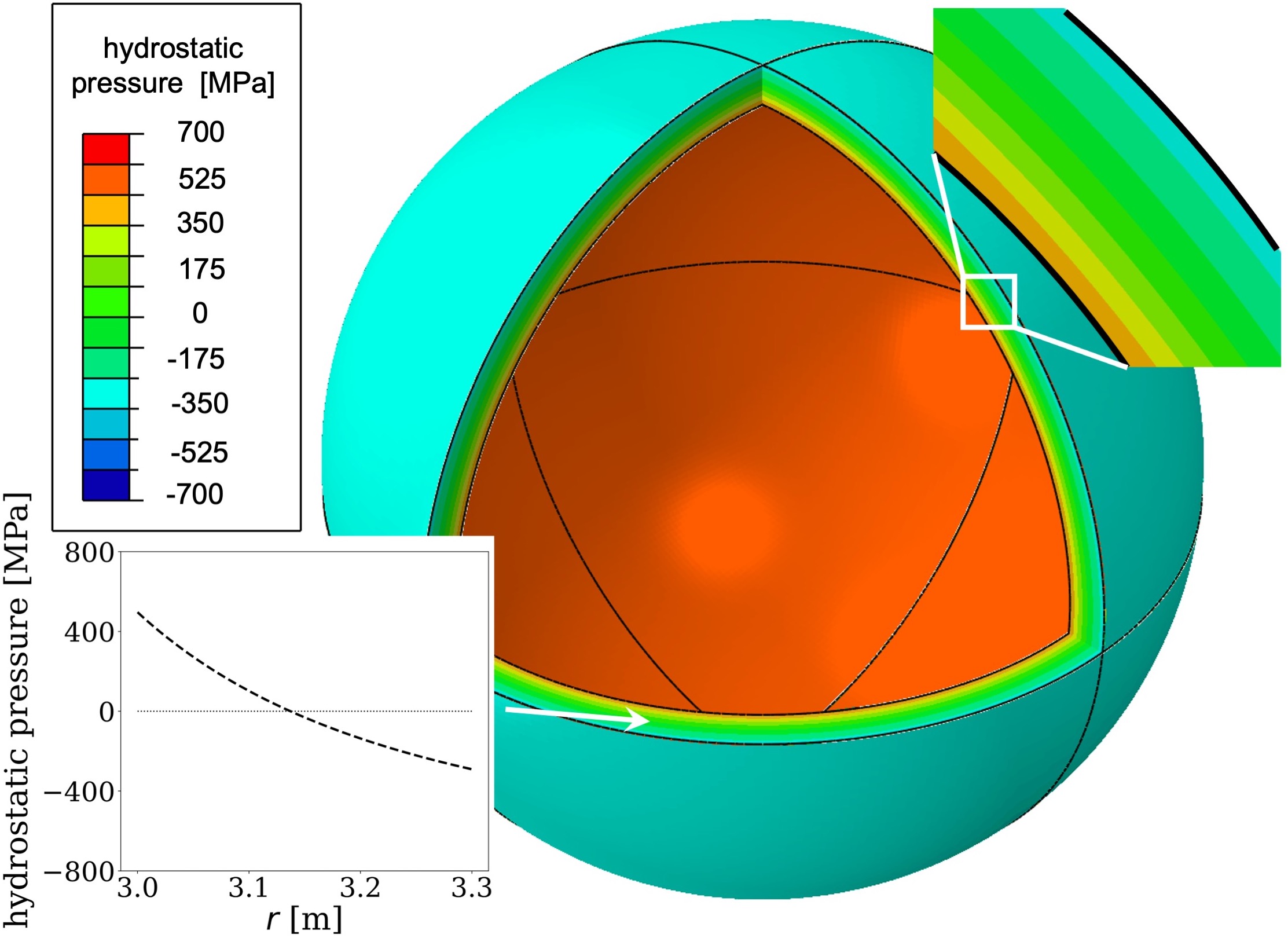}%
}\hfill
    \caption{Contour plots of the hydrostatic pressure induced by irradiation corresponding to the profiles of densities of relaxation volumes of defects $\omega(r)$ shown in Fig. \ref{fig:sph_omega} for case 1 in (a) and for case 2 in (b). The colour scale is the same in both plots, showing that stress is higher in case 1 than in case 2, despite the fact that the total relaxation volume of defects in the shell $\Omega _{tot}$ is the same in both cases (see Fig. \ref{fig:sph_omega}).}
    \label{fig:sphere_FEM}
\end{figure}

The circumferential (hoop) components $\sigma_{\theta\theta}(r)$ and $\sigma_{\phi\phi}(r)$ of the elastic stress tensor are
%
%
%
\begin{align}
    \sigma_{\theta\theta}(r) 
        &= {4\mu \over 9} \left({1+\nu \over 1-\nu}\right)
                \left[\wmean \left(1+{R_1^3\over 2r^3}\right)\right]\nonumber \\
        &+ {2\mu \over 3}\left({1+\nu \over 1-\nu}\right)
                \left[{1 \over r^3}\int \limits _{R_1}^r R^2\omega(R) \dint{R}\right]\nonumber \\
        &- {2\mu\over 3}\left({1+\nu \over 1-\nu}\right)\omega(r).
    \label{stress_ff}
\end{align}
Due to the symmetry of the problem, we have $\sigma_{\phi\phi}(r)=\sigma_{\theta\theta}(r)$, as can be readily confirmed by a direct calculation. Similarly to the radial component of elastic stress, both hoop components of stress vanish identically in the limit where $\omega(r)=const$ or, in other words, where defects are distributed spatially homogeneously throughout the volume of the shell.

The hydrostatic pressure developing in the shell as a result of accumulation of defects is
\begin{equation}
    p(r) = -{1\over 3}\sigma_{ii}(r) = -{1\over 3}\left[
                \sigma_{rr}(r) + \sigma_{\phi \phi}(r) + \sigma _{\theta\theta}(r)
            \right].
\end{equation}
Substituting expressions \eqref{stress_rr} and \eqref{stress_ff} in this equation, we find a surprisingly simple relation
\begin{equation}\label{pressure_spherical_shell}
    p(r) = {4\mu \over 9} \left( {1+\nu \over 1-\nu}\right)\left[\omega(r) - \wmean \right].
\end{equation}
Equation \eqref{pressure_spherical_shell} shows that pressure is positive where the local density of relaxation volumes is higher than its mean value, and negative where the local density of relaxation volumes is lower than the mean. The mean pressure, integrated over the entire volume of the shell, is zero.

For the purpose of illustrating the effect of different spatial distributions of defects in the volume of a component, as well as comparing the analytical and FEM solutions, we consider two functions $\omega(r)$ plotted in Fig.~\ref{fig:sph_omega}, where case 1 corresponds to equation \eqref{eq:om_sph_1} and case 2 corresponds to equation \eqref{eq:om_sph_2} in the appendix. The shell has the inner radius of $R_1=\SI{3.0}{m}$, the outer radius of $R_2=\SI{3.3}{m}$ and is assumed to be made of ferritic steel ($\mu=\SI{80}{\GPa}$, $\nu=0.29$). This is a simplified representation of the vacuum vessel of a fusion reactor. The FEM mesh used in simulations involves $1.33\times10^6$ hexahedral linear elements. No external stress is applied to the shell. 

High energy (\SI{14}{\MeV}) fusion neutrons attenuate in steels over distances of the order of tens of cm~\cite{Sato2003,Gilbert2013}. For example, in experiments~\cite{Greene_1969} only 10\% of the dose from a beam of \SI{14}{\MeV} neutrons reached 40~cm into steel and just 1\% of that transmitted dose was comprised of fast (\(\ge\)\SI{1}{\MeV} in energy) neutrons. Case 2 is representative of these conditions and assumes that $\omega(r)$ decays as $r^{-2}$ and that a proportion of the neutrons fully penetrate through the \SI{30}{cm} shell. Case 1, on the other hand, represents a more extreme example of neutrons moving through a highly attenuating material, causing so much damage that swelling saturates at shallow depths, and then being completely stopped within a distance of \(\sim\)\SI{20}{cm}. Radiation swelling reaches a dynamic saturation with the density of relaxation volumes of 2\% near the plasma, and drops to zero over the distance of approximately \SI{20}{cm}. The two profiles are chosen so that they have the same total relaxation volume of defects $\Omega_{tot}$.

The two distributions of $\omega(r)$ give rise to very different elastic stress patterns, illustrated in Fig. \ref{fig:sph_stresses}. In both cases, the elastic hoop stresses are of the order of 
\SI{1}{\GPa} and are higher than the radial component of elastic stress. The hoop stresses are negative close to the inner surface and positive near the outer surface of the shell. Intuitively this is clear since the part of the shell more exposed to irradiation attempts to expand, but is constrained by the less irradiated material, which in turn is under tension to achieve mechanical equilibrium. The maximum value of the von Mises stress is reached at the inner surface and is of the order of \SI{1015}{\MPa} in case 1 and \SI{745}{\MPa} in case 2. Radial displacement fields are also different due to the difference in $\omega(r)$, as shown in Fig. \ref{fig:sph_displ}. However, $u_r(r)$ takes exactly the same values at the inner and at the outer surfaces of the shell. This agrees with equation \eqref{swelling_surfaces} since $\Omega_{tot}$ is the same in case 1 as in case 2.

A comparison between Fig. \ref{fig:sph_stresses} and Fig. \ref{fig:sph_displ} illustrates an important point: a structure that upon a superficial external examination exhibits exactly the \emph{same} swelling, at least in terms of how its external surfaces move, can develop very different internal stresses. In the examples investigated here, the maximum hoop stress found in case 2 is less than half of that found in case 1.

Contour plots of the internal hydrostatic pressure are compared in Fig. \ref{fig:sphere_FEM}. In both cases, pressure is high and positive close to the inner surface of the shell, vanishes near the centre of the shell, and becomes negative towards the outer surface. The latter may have important implications for the stress-driven diffusion of interstitial elements such as hydrogen, helium or carbon, which would deplete the inner region and segregate towards the outer region of the component. Taking case 1 as an example, the equilibrium concentration of C at \SI{300}{\kelvin} would be about 30\% higher on the outside and about 25\% lower on the inside than in the unstressed condition.

\section{Finite element analysis of elastic stress in an irradiated tritium breeding blanket module}\label{sec:BB}

In this section we consider an example where the geometry of a component is too complex to admit an analytical solution, but which is not too dissimilar to what might be considered in the context of a design of a fusion power plant. We analyse, using FEM, the stress fields in a module of a fusion breeding blanket, assuming that it is exposed to a spatially varying flux of neutrons. 

\begin{figure}[t]
  \includegraphics[width=0.48\textwidth]{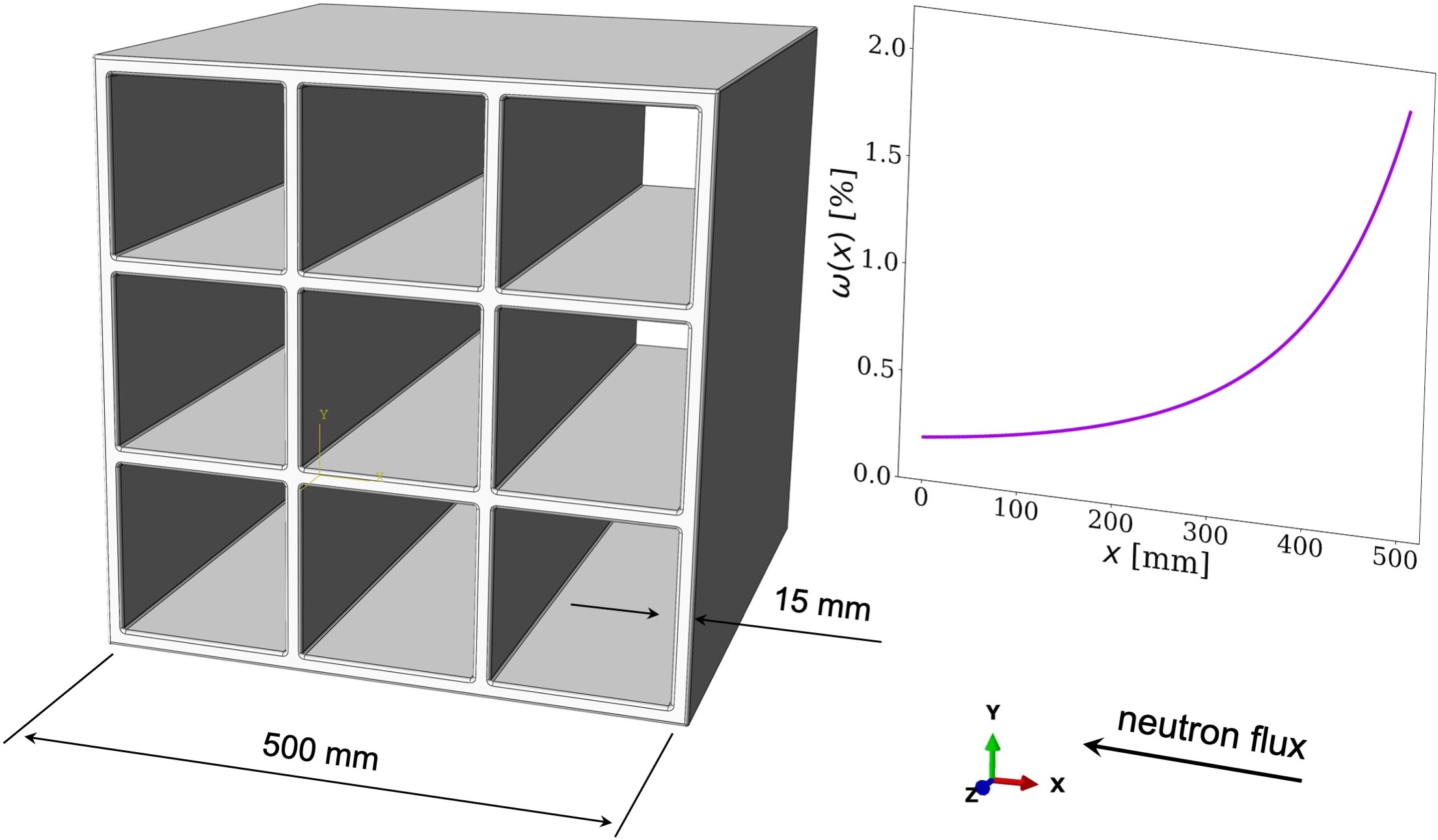}%
    \caption{Schematic sketch of a breeding blanket module. Neutron flux and the resulting swelling are assumed to depend only on coordinate $x$. The face of the component situated at $x=500$ mm is facing the plasma.}
    \label{fig:BB_geom}
\end{figure}

The breeding blanket is one of the most significant nuclear components of a fusion power plant, providing and enabling power extraction, tritium fuel sustainability, and radiation shielding \cite{Abdou2017FusEngDes,Federici2019FusEngDes}. A breeding blanket consists of individual modules containing tritium breeding materials and provides channels for coolant circulation. As an example, we consider a cubic structure with linear dimensions of \SI{500}{mm}, subdivided into 9 submodules. The wall thickness is assumed to be \SI{15}{mm}, and all the internal edges are filleted with a radius of \SI{3}{mm}. The structural material for the breeding blanket module is ferritic-martensitic steel Eurofer \cite{Rieth2008}. The structure is exposed to neutron irradiation from one of the sides, with the direction of the neutron flux being normal to the direction of flow of coolant. In this geometry, radiation exposure varies as a function of only one spatial coordinate \cite{Gilbert2013}. We assume that the maximum value of $\omega(x)$ is 2\% at the plasma-facing side of the module, see Fig. 7 of Ref. \cite{Zinkle2014}, and that it decreases down to 0.5\% across the component. The geometry of the module and the distribution of $\omega(x)$, represented by equation \eqref{eq:om_BB}, are shown in Fig. \ref{fig:BB_geom}.

The component is free of external loads, and is weakly constrained at the four corners of the face $z=0$. This is done to prevent its rigid motion but at the same time allows for its free deformation. $2.01\times10^6$ quadratic tetrahedral elements were used in numerical calculations, following the analysis of convergence given in Appendix~\ref{sec:convergence}.

\begin{figure}
\subfloat[\label{fig:BB_vM}]{%
  \includegraphics[width=0.4\textwidth]{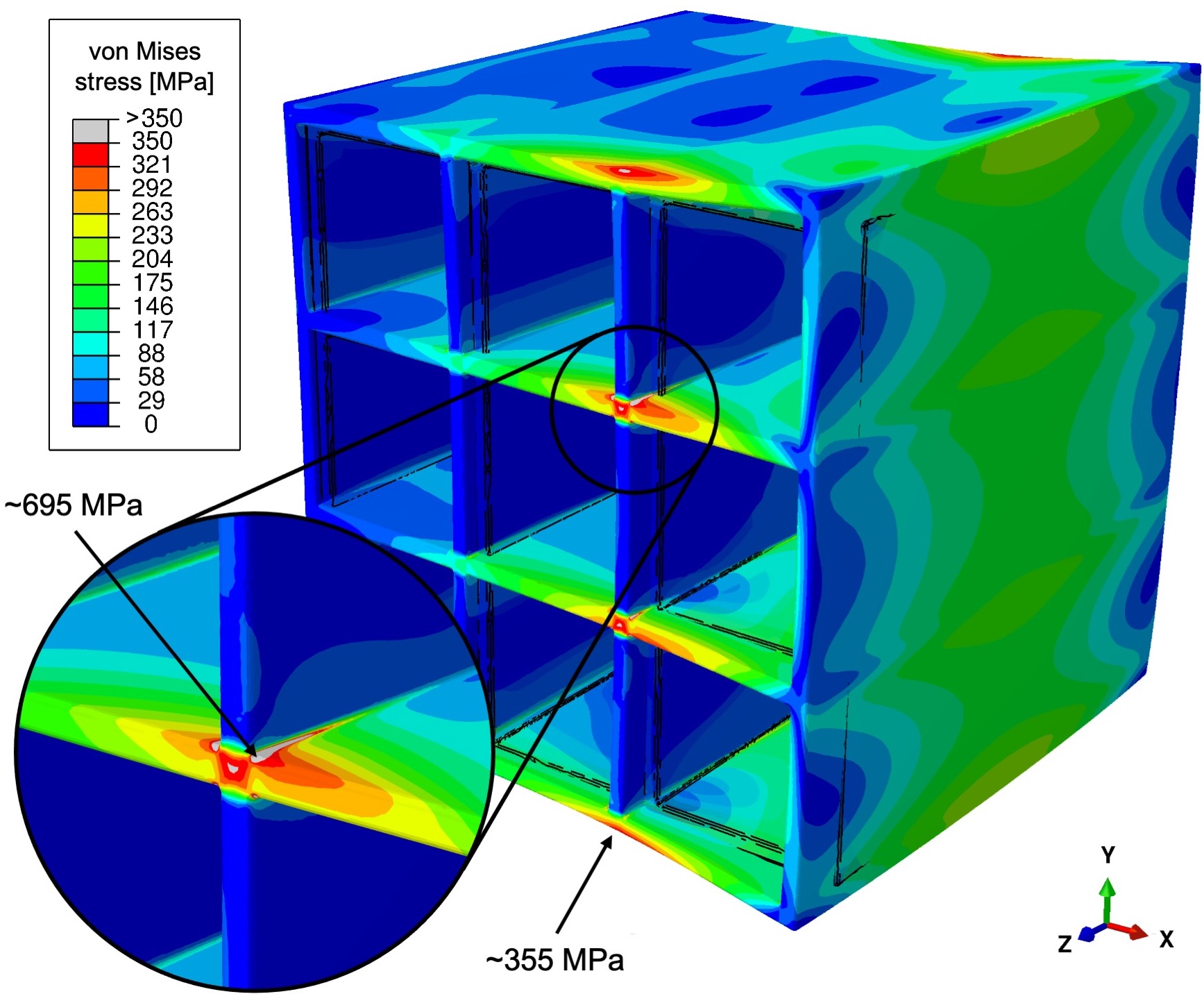}%
}\hfill
\subfloat[\label{fig:BB_MP}]{%
  \includegraphics[width=0.4\textwidth]{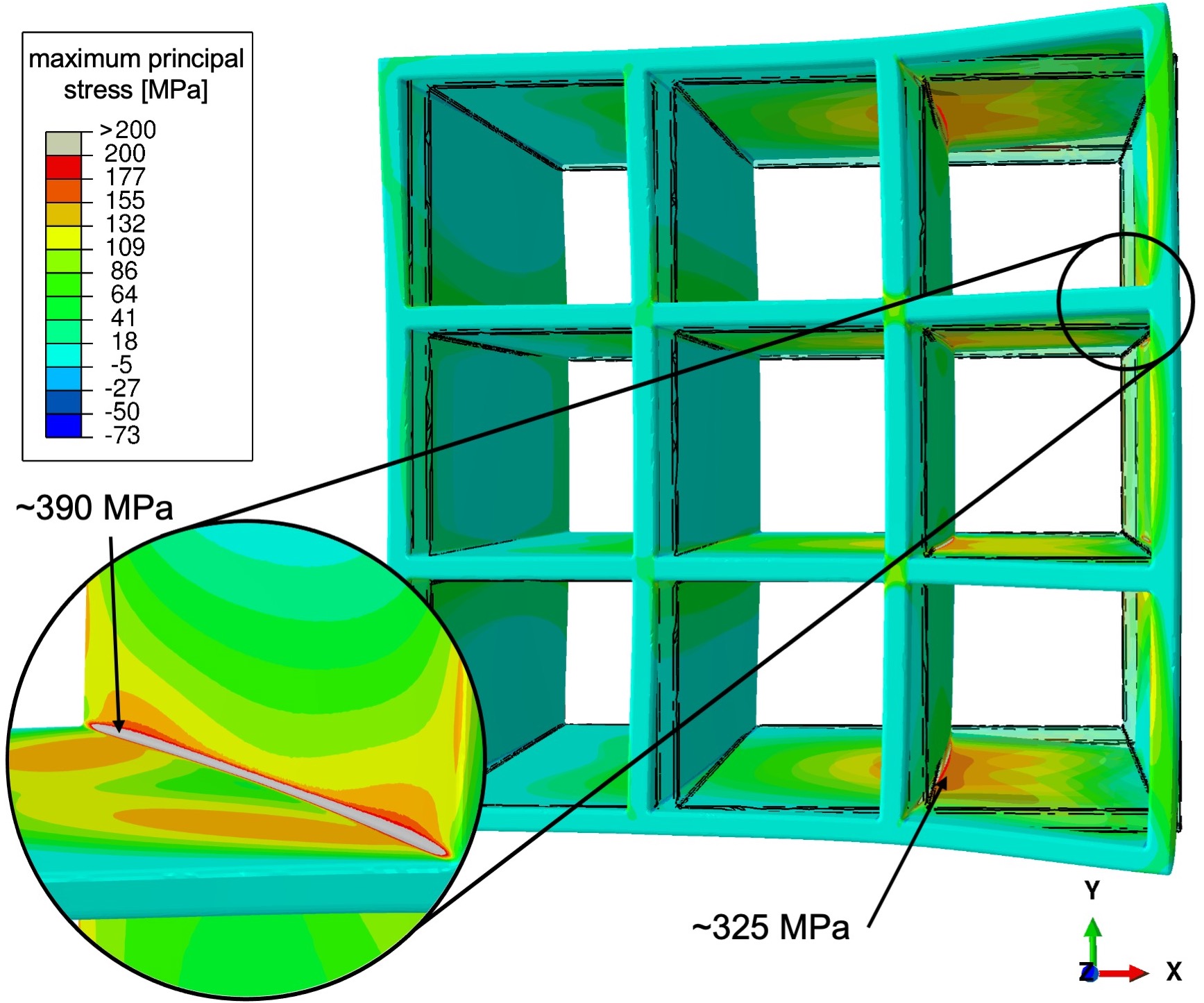}%
}\hfill
    \caption{Spatially varying exposure of materials to irradiation gives rise to stresses and distortions in the breeding blanket module, which in this example is assumed to be free from external geometric constraints and external load. (a) the local von Mises stress is maximum at the junction between the internal walls, where it is expected to exceed the yield stress of the material, whereas the maximum principal stress (b) is maximum at some of the junctions between  internal and external walls. The structure of the module before exposure to irradiation is outlined, and the scale of deformation is enlarged for clarity.}
    \label{fig:BB_stresses}
\end{figure}

For the given combination of component geometry and irradiation profile shown in Fig. \ref{fig:BB_stresses}, there are two locations where stresses are particularly high. Close to the points where internal walls intersect, the local von Mises stress reaches $\sim$\SI{695}{\MPa}. For comparison, the yield point of Eurofer is \SI{530}{\MPa} \cite{Fernandez2002}. At the junction between internal and external walls on the plasma-facing side, the maximum principal stress is close to $\sim$\SI{390}{\MPa}, suggesting that under irradiation it is this location in the breeding blanket module structure that would be susceptible to the nucleation of cracks.

\begin{figure}
\subfloat[\label{fig:BB_design2_vM}]{%
  \includegraphics[width=0.48\textwidth]{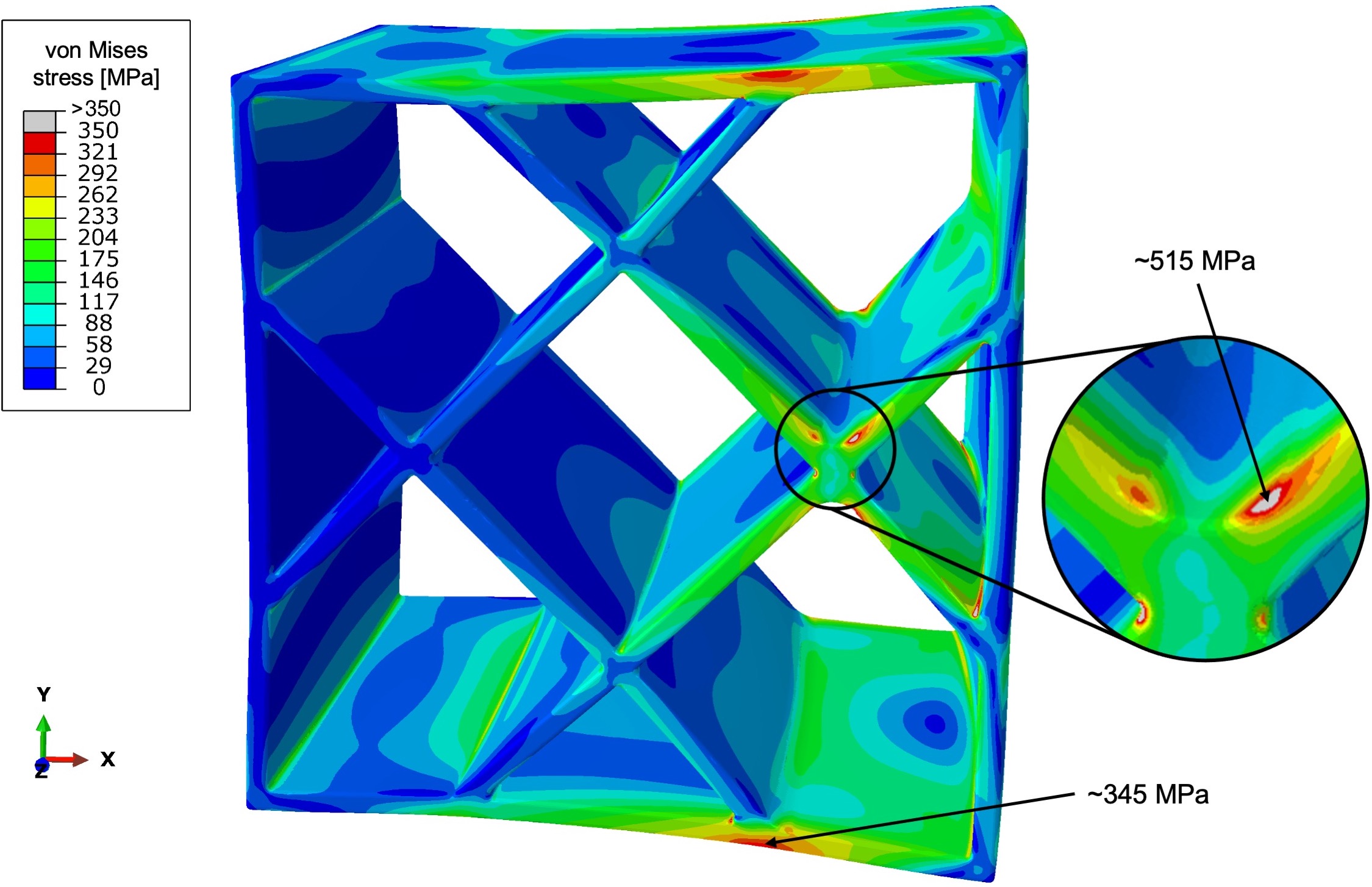}%
}\hfill
\subfloat[\label{fig:BB_design2_MP}]{%
  \includegraphics[width=0.48\textwidth]{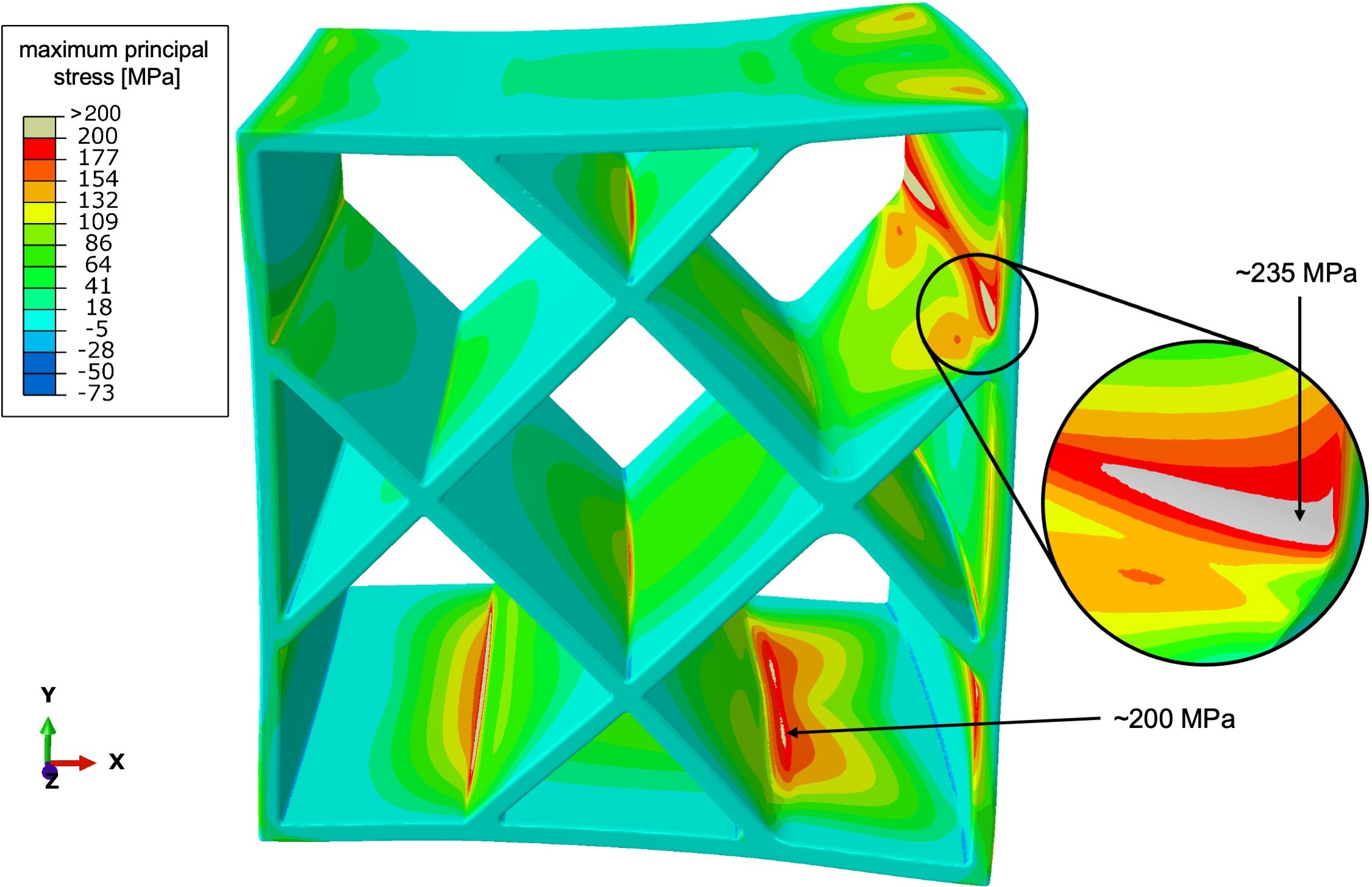}%
}\hfill
    \caption{Contour plots of (a) the local von Mises stress and (b) the maximum principal stress, computed for the modified blanket module design, to be compared with the initial design shown Fig. \ref{fig:BB_stresses}. The colour scale bars are the same in both figures. The number of locations where the stress is particularly high is reduced as a result of the rotation of the inner walls. Enlarging the radii of some of the fillets also reduces the magnitude of the local stress. In the zoomed regions, in comparison with the similar regions in Fig. \ref{fig:BB_stresses}, the von Mises stress is now lower by 25\%, and the maximum principal stress is lower by 40\%. The scale of deformation of the structure is magnified for clarity. The initial undeformed structure is not shown.}
    \label{fig:BB_design2}
\end{figure}

From this analysis, and the results given in the preceding sections of this study, we find that the effect of irradiation-induced swelling should be taken into account already at the stage of component design. To exemplify this, we start from the initial design shown in Fig. \ref{fig:BB_geom} and apply two adjustments illustrated in Fig. \ref{fig:BB_design2}. First, the internal walls can be rotated by 45$^{\circ}$ while keeping their length constant. This reduces the number of internal junctions near the plasma-facing side of the component from two to one. Second, the radius of the fillets can be increased where the stresses due to irradiation are expected to be higher. To explore the effect of structural changes, the radius of fillets in the most stressed internal junction was increased to \SI{20}{mm}, whereas the radius of the four fillets near the plasma-facing side of the component was increased to \SI{30}{mm}. This produces a modified design, which we subsequently tested using FEM numerical analysis. The von Mises and maximum principal stresses developing in the second design, under the same irradiation conditions, are shown in Fig. \ref{fig:BB_design2}, to be compared with Fig. \ref{fig:BB_stresses}. The FEM model involved $2.11\times10^6$ elements to ensure convergence. High stresses arise at locations that are similar to those of the first design, but the local von Mises and maximum principal stresses are now lower by 25\% and 40\%, respectively. The highest maximum principal stress has now developed at other locations in the structure. Further modifications of the design can now be explored in the framework of iterative numerical design optimisation, which is beyond the scope of this work.

We observe that irradiation gives rise to sizeable deformations of the component structure. Given the distribution of relaxation volumes of defects, the module expands by approximately \SI{2.5}{mm} in the $x$, \SI{3.3}{mm} in the $y$ and \SI{3.1}{mm} in the $z$ direction. This expansion is the same for the two design variants considered above, and is close to the deformation of the same structure not containing any inner walls under the same irradiation conditions. The modified module expands by 3.1, 3.4 and \SI{3.3}{mm} in the $x$, $y$ and $z$ directions, respectively. 

What we find is different from what is known in conventional structural engineering where deformation of a structure subjected to an external load can be reduced by means of internal reinforcements. This methodology does not apply here, since the reinforcements themselves expand following their exposure to irradiation.

Our final note in this section refers to the distribution of hydrostatic pressure $p({\bf x})$ in an irradiated component. Equation \eqref{pressure_spherical_shell} shows that in a spherical shell able to expand freely, where both surfaces are subject to the traction-free boundary conditions, $p({\bf x})$ vanishes exactly after averaging over the entire volume of the component. The same applies to the case of a breeding blanket module, provided that it is free of external constraints. The volume-average $p({\bf x})$, evaluated numerically using all the integration points of the FEM model, is of the order of \SI{1}{\MPa} for the design shown in Fig.  \ref{fig:BB_stresses} and \SI{2}{\MPa} for the modified design shown in Fig. \ref{fig:BB_design2}. These values are very small in comparison with the maximum and minimum values of pressure of \SI{285}{\MPa} and \SI{-172}{\MPa} computed for the initial component design, and \SI{208}{\MPa} and \SI{-139}{\MPa} computed for the modified design, respectively. 

The above analysis shows that in the breeding blanket as well as in some other cases that we studied, the yield point of the material was locally exceeded for an assumed swelling profile. To model the resulting response of the structure, the evolution of an irradiated material has to be modelled using elasto-plastic constitutive rules in addition to the laws describing the generation of defects by irradiation. In this context, we expect that the irradiation-driven stress relaxation \cite{Peacock2004,Luzginova2011} is going to occur even at relatively lower stresses, partially alleviating them but resulting in the even greater deformation of irradiated components and thus potentially generating stresses in other parts of the reactor structure. Modelling the direct stress-induced plasticity and the slower  irradiation-driven stress relaxation, as well as taking into account the temperature and stress dependence of $\omega _{ij}({\bf x})$ defines a broad scope for future work, extending beyond this study. We note that, irrespectively of the occurrence of stress relaxation, the relations between the relaxation volume density $\omega _{ij}({\bf x})$ and elastic stress and strain established in this study remain unchanged, since stress relaxation only affects the form of $\omega _{ij}({\bf x})$ and not the elastic field that it generates. 

\section{conclusions}

In this study, we developed a method for computing elastic stress and strain fields developing in structural components of a fusion power plant exposed to irradiation (or indeed any other nuclear facility where components are exposed to a significant irradiation fluence). The approach is based on a {\it defect eigenstrain theorem} (eq. \ref{defect_eigenstrain_theorem}) that states that the spatially varying density of relaxation volume tensors of defects produced by neutron irradiation (eq. \ref{volume_tensor_density_ensemble}), which is a quantity that can be computed from microscopic considerations \cite{MaPRM2019a,MaPRM2019b,Mason2019,MaPRM2020,MaPRM2021}, is identical to the spatially varying field of eigenstrain \cite{Mura}.  

In the absence of external geometric constraints or, equivalently, under the traction-free boundary conditions, the elastic stress and internal pressure generated by irradiation always vanish after integration over the entire volume of a component. As a result, a spatially varying exposure of a component to irradiation invariably produces a spatially varying field of stress, compressive at locations where the density of defects is higher, and tensile at locations where the density of defects is lower, see e.g. equation (\ref{pressure_spherical_shell}). The occurrence of irradiation embrittlement \cite{vanderSchaaf2009} makes operating critical parts of a component under compressive stress preferable to operating under the tensile, negative pressure, conditions.    

The magnitude of elastic stress developing in a component as a result of irradiation can be fairly high. It can be estimated as $\sigma ({\bf x})\sim \mu [\overline \omega -\omega ({\bf x})]$, where $\mu $ is the shear modulus of the material and $\overline \omega - \omega ( {\bf x})$ is a measure of deviation of the spatially varying density of relaxation volumes of defects from its average value. The density of relaxation volumes of defects is a dimensionless quantity and, depending on the experimental conditions, it varies from a fraction of a percent \cite{Mason2020} to several percent \cite{Zinkle2014,Garner2020}. Bearing in mind that for structural steels and tungsten $\mu $ is close to a hundred GPa, the irradiation-induced stress can reach several hundred MPa, as illustrated in Fig. \ref{fig:BB_design2}. At the same time, a spatially homogeneous distribution of defects does not generate any stress in a free body, although it could still produce large deformations. High stresses can still develop if deformations are constrained.

There are now firm indications suggesting that stress developing in a component as a result of irradiation can elastically polarise defects, resulting in a highly non-linear response of microstructure to stress {\it and} irradiation. While this does not alter the fundamental link, equations \eqref{elastic_strain} -- \eqref{effective_body_force}, between the eigenstrain of the defects and the elastic stress and strain fields that they produce, the combined non-linear self-action of stress and high temperature on the microstructure, already noted in recent simulations and observations \cite{Chen2010,Boleininger2019,Mason2020,Lai2020}, requires extensive analysis.  

Concluding this study, we point out that the link between the microscopic properties of defects and the macroscopic stress and strain fields that they produce in reactor components, provides an example of a complete multiscale study involving a full treatment of microscopic and macroscopic scales and enabling the assessment of performance of materials and components under realistic operating conditions that can now be quantitatively modelled using advanced supercomputer facilities.



\begin{acknowledgments}
We thank I. Chapman for the provision of laboratory facilities, and are grateful to R. Akers, P.-W. Ma, A. Davis, A. London, F. Hofmann, G. Pintsuk and M. Rieth for helpful discussions. SLD would like to thank A. P. Sutton for a detailed discussion of scientific analysis by J. D. Eshelby \cite{Eshelby1956}, which stimulated this study. This work has been carried out within the framework of the EUROfusion Consortium and has received funding from the Euratom research and training programme 2014-2018 and 2019-2020 under grant agreement No. 633053 and from the RCUK Energy Programme, grant No. EP/T012250/1. To obtain further information on the data and models underlying the paper please contact PublicationsManager@ukaea.uk. The views and opinions expressed herein do not necessarily reflect those of the European Commission. We gratefully acknowledge the provision of computing resources by the IRIS (STFC) Consortium.
\end{acknowledgments}

\begin{appendix}

\section{Deriving the relaxation volume}\label{sec:A:relaxation}

In the following derivation, subscripts after a comma denote differentiation ($f_{,l} = \partial f/\partial x_l$), with primed subscripts referring to differentiation with respect to the primed variable  ($f_{,l'} = \partial f/\partial x_l'$).

Consider an elastic body with the volume defined by region $V$ with a surface $S = \partial V$ free of tractions. Mura \cite{Mura} defines the total strain $\epsilon_{ij}^{(tot)}$ as the sum of elastic strain $\epsilon_{ij}$ and eigenstrain $\epsilon_{ij}^*$:
\begin{equation}
\epsilon_{ij}^{(tot)} = \epsilon_{ij} + \epsilon_{ij}^*,
\end{equation}
with the total strain being compatible $\epsilon_{ij}^{(tot)} = \frac{1}{2}(u_{i,j} + u_{j,i})$. Similarly, we may introduce the eigenstress $\sigma_{ij}^*$ as a quantity related to eigenstrain $\epsilon_{ij}^*$ through Hooke's law:
\begin{equation}
\sigma_{ij}^* = C_{ijkl} \epsilon_{kl}^* .
\end{equation}
The Cauchy stress, or elastic stress, is then given by
\begin{equation}
\sigma_{ij} = C_{ijkl} \epsilon_{kl} = C_{ijkl}\left( \epsilon_{kl}^{(tot)} - \epsilon_{kl}^*\right),
\end{equation}
from which follows the equilibrium condition inside the body $\vec{x} \in V$:
\begin{equation}
\sigma_{ij,j} = C_{ijkl}\left( \epsilon_{kl,j}^{(tot)} - \epsilon_{kl,j}^*\right) = 0.
\end{equation}
The above expression yields a relation between the body force and the eigenstrain: 
\begin{equation}\label{eq:A:bodyforce}
C_{ijkl} \epsilon_{kl,j}^{(tot)} = C_{ijkl} \epsilon_{kl,j}^*  = \sigma_{ij,j}^* \equiv -f_i.
\end{equation}
Equivalently, for a body free of external surface forces, the equilibrium condition at the surface $\vec{x} \in S$ follows as
\begin{equation}\label{eq:A:traction}
C_{ijkl} \epsilon_{kl}^{(tot)} n_j = C_{ijkl} \epsilon_{kl}^* n_j  = \sigma_{ij}^* n_j \equiv t_i,
\end{equation}
where $n_j(\vec{x})$ is the outwards pointing normal vector of surface $S$ at point $\vec{x} \in S$.

The displacement field in a body subjected to a body-force is given by [Ref. \cite{Sutton2020}, Eq.~(4.12)]
\begin{equation}\label{eq:A:displacement1}
\begin{aligned}
u_j(\vec{x}) =  &\hphantom{+}   \int_V G_{jk}(\vec{x}-\vec{x}') f_k(\vec{x}') \dint{^3 x'} \\
                &+\int_S        G_{jk}(\vec{x}-\vec{x}') t_k(\vec{x}') \dint{^2 x'} \\
                &-\int_S        G_{jk,p'}(\vec{x}-\vec{x}') C_{kpmi} u_i(\vec{x}') n_m \dint{^2 x'},
\end{aligned}
\end{equation}
where $f_k$ is the body force and $t_k = C_{kpmi} u_{i,m'} n_p$ is the surface traction.

Now consider a case where the only acting body and surface forces are caused by eigenstrain. We insert the definitions of body force \eqref{eq:A:bodyforce} and traction \eqref{eq:A:traction} into \eqref{eq:A:displacement1}, arriving at:
\begin{equation}\label{eq:A:displacement2}
\begin{aligned}
u_j(\vec{x}) =  &-\int_V G_{jk}(\vec{x}-\vec{x}') C_{kpmi} \epsilon_{mi,p'}^*(\vec{x}') \dint{^3 x'} \\
                &+\int_S G_{jk}(\vec{x}-\vec{x}') C_{kpmi} \epsilon_{mi}^*(\vec{x}') n_p \dint{^2 x'}\\
                &-\int_S G_{jk,p'}(\vec{x}-\vec{x}') C_{kpmi} u_i(\vec{x}') n_m \dint{^2 x'}.
\end{aligned}
\end{equation}
We first apply the product rule for differentiation to the first term in \eqref{eq:A:displacement2}, noting that $G_{jk,p'} = -G_{jk,p}$,
\begin{equation}
\begin{aligned}
 &-\int_V G_{jk}(\vec{x}-\vec{x}') C_{kpmi} \epsilon_{mi,p'}^*(\vec{x}') \dint{^3 x'} \\
=&-\int_V \left[G_{jk}(\vec{x}-\vec{x}') C_{kpmi} \epsilon_{mi}^*(\vec{x}')\right]_{,p'} \dint{^3 x'} \\
 &-\int_V G_{jk,p}(\vec{x}-\vec{x}') C_{kpmi} \epsilon_{mi}^*(\vec{x}') \dint{^3 x'},
\end{aligned}
\end{equation}
and then proceed with applying the divergence theorem, to obtain 
\begin{equation}\label{eq:A:proddiv}
\begin{aligned}
 &-\int_V  G_{jk}(\vec{x}-\vec{x}') C_{kpmi} \epsilon_{mi,p'}^*(\vec{x}') \dint{^3 x'}\\
=&-\int_S  G_{jk}(\vec{x}-\vec{x}') C_{kpmi} \epsilon_{mi}^*(\vec{x}') n_p \dint{^2 x'}\\
 &-\int_V  G_{jk,p}(\vec{x}-\vec{x}') C_{kpmi} \epsilon_{mi}^*(\vec{x}') \dint{^3 x'}.
\end{aligned}
\end{equation}
Substituting \eqref{eq:A:proddiv} into \eqref{eq:A:displacement2}, we arrive at
\begin{equation}\label{eq:A:displacement3}
\begin{aligned}
u_j(\vec{x}) =  &-\int_V G_{jk,p}(\vec{x}-\vec{x}') C_{kpmi} \epsilon_{mi}^*(\vec{x}') \dint{^3 x'}\\
                &+\int_S G_{jk,p}(\vec{x}-\vec{x}') C_{kpmi} u_i(\vec{x}') n_m \dint{^2 x'}.
\end{aligned}
\end{equation}
In preparation of the next step, it is helpful to express the second term in \eqref{eq:A:displacement3} as a volume integral by using the divergence theorem, noting that $G_{jk,pm'} = -G_{jk,pm}$,
\begin{equation}\label{eq:A:proddiv2}
\begin{aligned}
            & \int_S G_{jk,p}(\vec{x}-\vec{x}') C_{kpmi} u_i(\vec{x}') n_m \dint{^2 x'}\\
=\phantom{-}& \int_V \left[G_{jk,p}(\vec{x}-\vec{x}') C_{kpmi} u_i(\vec{x}')\right]_{,m'} \dint{^3 x'}\\
=          -& \int_V G_{jk,pm}(\vec{x}-\vec{x}') C_{kpmi} u_i(\vec{x}')  \dint{^3 x'}\\
            & \int_V G_{jk,p}(\vec{x}-\vec{x}') C_{kpmi} u_{i,m'}(\vec{x}') \dint{^3 x'}.
\end{aligned}
\end{equation}
Substituting \eqref{eq:A:proddiv2} into \eqref{eq:A:displacement3}, we find
\begin{equation}\label{eq:A:displacement4}
\begin{aligned}
u_j(\vec{x}) =  &-\int_V G_{jk,p}(\vec{x}-\vec{x}') C_{kpmi} \epsilon_{mi}^*(\vec{x}') \dint{^3 x'} \\
                &-\int_V G_{jk,pm}(\vec{x}-\vec{x}') C_{kpmi} u_i(\vec{x}') \dint{^3 x'}\\
                &+\int_V G_{jk,p}(\vec{x}-\vec{x}') C_{kpmi} u_{i,m'}(\vec{x}') \dint{^3 x'}.
\end{aligned}
\end{equation}

Defining the volume-averaged distortion tensor $\overline{u}_{j,l}$ by the relation
\begin{equation}\label{eq:A:distdef}
V \overline{u}_{j,l} =  \int_V \, u_{j,l}(\vec{x})\dint{^3 x},
\end{equation}
we substitute \eqref{eq:A:displacement4} into \eqref{eq:A:distdef} and write the volume integral over $\dint{^3 x'}$ as a convolution
\begin{equation}\label{eq:A:relaxtensor1}
\begin{aligned}
V \overline{u}_{j,l}  =  &-\int_V 
                    C_{kpmi} \left(G_{jk,pl}  \ast \epsilon_{mi}^* \right)(\vec{x})\, \dint{^3 x} \\
                         &-\int_V C_{kpmi} \left(G_{jk,pml} \ast u_i\right)(\vec{x})\, \dint{^3 x} \\
                         &+\int_V C_{kpmi} \left(G_{jk,pl}  \ast u_{i,m'}\right)(\vec{x})\, \dint{^3 x}.
\end{aligned}
\end{equation}
We proceed with simplifying the second line in \eqref{eq:A:relaxtensor1}. From the definition of the elastic Green's function, we know that [Sutton \cite{Sutton2020}, Eq.~(4.8)]
\begin{equation}\label{eq:A:greensdirac}
C_{kpim} G_{ij,mp}(\vec{x}-\vec{x}') = -\delta_{jk} \delta(\vec{x}-\vec{x}'),
\end{equation}
which, after manipulation of indices and making use of the symmetries of the stiffness tensor and the Green's function, is equal to
\begin{equation}\label{eq:A:greensdirac2}
C_{kpmi} G_{jk,mp}(\vec{x}-\vec{x}') = -\delta_{ji} \delta(\vec{x}-\vec{x}').
\end{equation}
The substitution of \eqref{eq:A:greensdirac2} into the second line of \eqref{eq:A:relaxtensor1} yields
\begin{equation}
\begin{aligned}
\phantom{=}-&   \int_V  C_{kpmi} \left(G_{jk,pml} \ast u_i\right)(\vec{x})\, \dint{^3 x} \\
= -&            \int_V \int_V \left[
            -\delta_{ji} \delta_{,l}(\vec{x}-\vec{x}')\right] u_i(\vec{x}')  \dint{^3 x'}  \dint{^3 x}\\
= &            \int_V u_{j,l}(\vec{x}) \dint{^3 x} =  V \overline{u}_{j,l},
\end{aligned}
\end{equation}
where we used $\partial_x (\delta \ast f)(x) = \partial_x f(x) = f'(x)$.
Cancelling out $V \overline{u}_{j,l}$ in \eqref{eq:A:relaxtensor1} hence leads to 
\begin{equation}\label{eq:A:relaxtensor2}
\begin{aligned}
    &\phantom{=}    \int_V  C_{kpmi} \left(G_{jk,pl}  \ast \epsilon_{mi}^* \right)(\vec{x})\, \dint{^3 x}\\
    &=              \int_V  C_{kpmi} \left(G_{jk,pl}  \ast u_{i,m'}\right)(\vec{x})\, \dint{^3 x}.
\end{aligned}
\end{equation}
Next, substituting 
\begin{equation}
C_{kpmi} u_{i,m'} = C_{kpmi} \epsilon_{im}^{(tot)} = C_{kpmi} (\epsilon_{im} + \epsilon_{im}^*)
\end{equation}
into the second line of \eqref{eq:A:relaxtensor2} and simplifying the resulting expression, we arrive at 
\begin{equation}\label{eq:A:relaxtensor3}
0   =   \int_V C_{kpmi} \left(G_{jk,pl}  \ast \epsilon_{im} \right)(\vec{x})\, \dint{^3 x}.
\end{equation}
The above expression is in effect a volume integral over a convolution between two functions $f$ and $g$, which may also be expressed as
\begin{equation}\label{eq:A:convint2}
\int_V \left(f \ast g\right)(\vec{x})\, \dint{^3 x}
= \left(\int_V  f(\vec{x}) \dint{^3 x} \right) \left(\int_V g(\vec{x}) \dint{^3 x} \right).
\end{equation}
Applying Eq.~\eqref{eq:A:convint2} to Eq.~\eqref{eq:A:relaxtensor3}, we arrive at 
\begin{equation}\label{eq:A:relaxtensor4}
0   =    C_{kpmi} \left[\int_V G_{jk,pl}(\vec{x}) \dint{^3 x} \right] V \overline{\epsilon}_{im},
\end{equation}
where $\overline{e}_{im} = V^{-1}\int \epsilon_{im}(\vec{x}) \dint{^3 x}$ is the volume-averaged elastic strain tensor. 

The so-called \textit{auxiliary tensor}
\begin{equation}
D_{jkpl} =   \int_V G_{jk,pl}(\vec{x}) \dint{^3 x}
\end{equation}
is not generally zero, enabling us to conclude that the volume-averaged elastic strain tensor vanishes
\begin{equation}
 \overline{\epsilon}_{im} = V^{-1}\int_V \epsilon_{im}(\vec{x})  \dint{^3 x} = 0.
\end{equation}
Note that the same applies to the volume-averaged elastic stress tensor, $\overline{\sigma}_{im} = 0$.

Finally, recalling the definition of the volume relaxation tensor
\begin{equation}
    \Omega_{ij} = \int_V \epsilon_{ij}^{(tot)}(\vec{x}) \dint{^3 x} = V \overline{\epsilon}^{(tot)}_{ij} = V \overline{\epsilon}_{ij} + V \overline{\epsilon}_{ij}^*,
\end{equation}
and substituting $\overline{\epsilon}_{ij} = 0$, we arrive at the central result
\begin{equation}\label{eq:A:relaxtensorfinal}
    \Omega_{ij} = V \overline{\epsilon}_{ij}^*.
\end{equation}

To conclude, the elastic field $\epsilon_{ij}$ does not contribute to volume change in a body with the surface free of tractions. Any volume change is solely effected by the volume-average eigenstrain field $\epsilon_{ij}^*$, as shown by expression \eqref{eq:A:relaxtensorfinal}. The proof above is general insofar as the eigenstrain field may be non-zero across the whole of the body, including the surface.

We can also recover the well known relation between relaxation volume tensor and dipole tensor \cite{PRB2018,NF2018}
\begin{equation}
    \Omega_{ij} = V \overline{\epsilon}_{ij}^* = S_{ijkl} C_{klmn} V \overline{\epsilon}_{mn}^* = S_{ijkl} P_{kl},
\end{equation}
where $P_{kl} = V \overline{\sigma}_{kl}^*$ is the dipole tensor.

\section{Volume changes in linear elasticity}\label{sec:A:volumechange}

Consider a body that underwent deformation such that point $\vec{x}$ in the initial configuration is at $\vec{x}'$ in the deformed configuration. The initial and deformed coordinates are related through the displacement field
\begin{equation}
\vec{x}' = \vec{x}+\vec{u}(\vec{x}).
\end{equation}
Considering the displacement as a coordinate transformation from $\vec{x}\rightarrow \vec{x}'$, the volume of the body is given by the volume integral
\begin{equation}
V = \int_V \mathrm{det}(\mathbf{J})\dint{x}\dint{y}\dint{z},
\end{equation}
where $\mathrm{det}(\mathbf{J})$ is the determinant of the Jacobian matrix
\begin{equation}
J_{ij} = \frac{\partial x_i'}{\partial x_j} = \delta_{ij} + u_{i,j}(\vec{x}),
\end{equation}
given by
\begin{equation}
\begin{aligned}
\mathrm{det}(\mathbf{J}) &= 1 + u_{i,i} + \frac{1}{2}\left(u_{i,i}u_{j,j} - u_{i,j}u_{j,i}\right) \\
                         &\phantom{=} + u_{1,i} u_{2,j} u_{3,k} \epsilon_{ijk},
\end{aligned}
\end{equation}
where $\epsilon_{ijk}$ is the Levi-Civita tensor.

In the limit of small strain
\begin{equation}
\mathrm{det}(\mathbf{J}) \approx 1 + u_{i,i},
\end{equation}
leading to the expression for the volume of the body \cite{LandauElasticity}
\begin{equation}
V = \int_V \left(1+u_{i,i}\right)\dint{x}\dint{y}\dint{z}.
\end{equation}
With this in mind, even if the displacement field has been obtained using linear elasticity theory, expression
\begin{equation}
\Omega _{tot}= \int_V \epsilon_{ii}^{(tot)} ({\bf x})\dint{V} 
\end{equation}
only captures the volume change of the distorted body to first order in strain. This is typically a valid approximation as mean strains in linear elasticity theory are usually limited to a few percent. 

\section{Relaxation volume density profiles}

In the treatment of distortions in a rectangular foil considered in Sec. \ref{sec:rectangle}, function $\omega(z)$ was taken to be proportional to the ion implantation profile. Three tungsten ion implantation profiles were considered in Sec. \ref{sec:rectangle}, corresponding to ion energies of 10, 20 and 50 MeV. The data points that were generated using the SRIM package~\cite{srim2008,ZIEGLER20101818} are shown in Fig. \ref{fig:dpa_profile_film}. These three data sets were fitted using the function
\begin{equation}\label{eq:fit_dpa}
    f(\zeta)=\frac{c_1+c_2\zeta+c_3\zeta^2}{\left[1+\text{exp}\left(\frac{\zeta-c_4}{c_5}\right)\right]^{c_6}}
\end{equation}
where $\zeta=h-z$ is the depth variable used in Fig. \ref{fig:dpa_profile_film}. Using the least squares regression we determined the fitting parameters given in table~\ref{tab:fittingpara}.

\begin{table}[t]
\caption{Parameters used in equation \eqref{eq:fit_dpa} for fitting the data plotted in Fig. \ref{fig:dpa_profile_film} and used in FEM simulations described in Sec. \ref{sec:rectangle}.}
\label{tab:fittingpara}
\begin{ruledtabular}  
\begin{tabular}{lccc}
    ion energy [MeV] & 10 & 20 & 50 \\\hline
    $c_1$ [-] & 0.2016 & 0.1275 & 0.0705 \\
    $c_2$ [\textmu m$^{-1}$] & 0.2177 & 0.0568 & -0.0022 \\
    $c_3$ [\textmu m$^{-2}$]  & 0.4401 & 0.0958 & 0.0253 \\
    $c_4$ [\textmu m] & 1.1385 & 2.0470 & 3.6592 \\
    $c_5$ [\textmu m] & 0.1967 & 0.2504 & 0.2444 \\
    $c_6$ [-] & 5.6681 & 4.4973 & 2.1856 \\
  	\end{tabular}
\end{ruledtabular}
\end{table}

Function $\omega(\zeta)$ was then assumed to be proportional to $f(\zeta)$. The proportionality constant was set to 0.03437 so that the peak of the 20 MeV curve corresponds to the relaxation volume density of 1\%.

For the cylindrical tube case explored in Sec. \ref{sec:cylinder}, the profile of $\omega(\rho)$ plotted in Fig. \ref{fig:omega_cyl} is
\begin{equation}\label{eq:om_cyl_FEM}
    \omega(\rho)=\frac{0.01}{(4.0-\rho)^2},
\end{equation}
where $2.5<\rho<3.0$ is in mm.

For the case of a spherical shell described in Sec. \ref{sec:sphere}, the profiles of $\omega(r)$ plotted in Fig. \ref{fig:sph_omega} are
\begin{equation}\label{eq:om_sph_1}
    \omega(r)=\frac{0.02}{1+\text{exp}\left(\frac{r-3.15}{0.01}\right)}
\end{equation}
for case 1 and
\begin{equation}\label{eq:om_sph_2}
    \omega(r)=\frac{0.005}{(r-2.5)^2}-0.0027843
\end{equation}
for case 2. Both functions are defined on the interval $3.0<r<3.3$, where $r$ is given in meter units. The second term in equation \eqref{eq:om_sph_2} ensures that integral $\int_V\omega ({\bf x})\text{d}V$ has the same value in both cases.

For the breeding blanket module of Sec. \ref{sec:BB}, function $\omega(x)$ is defined as
\begin{equation}\label{eq:om_BB}
    \omega(x)=\frac{1069}{(x-731)^2}
\end{equation}
where $0<x<500$ is expressed in mm units.

\section{FEM convergence analysis}\label{sec:convergence}
\begin{figure}
\subfloat[\label{fig:convergence_1}]{%
  \includegraphics[width=0.48\textwidth]{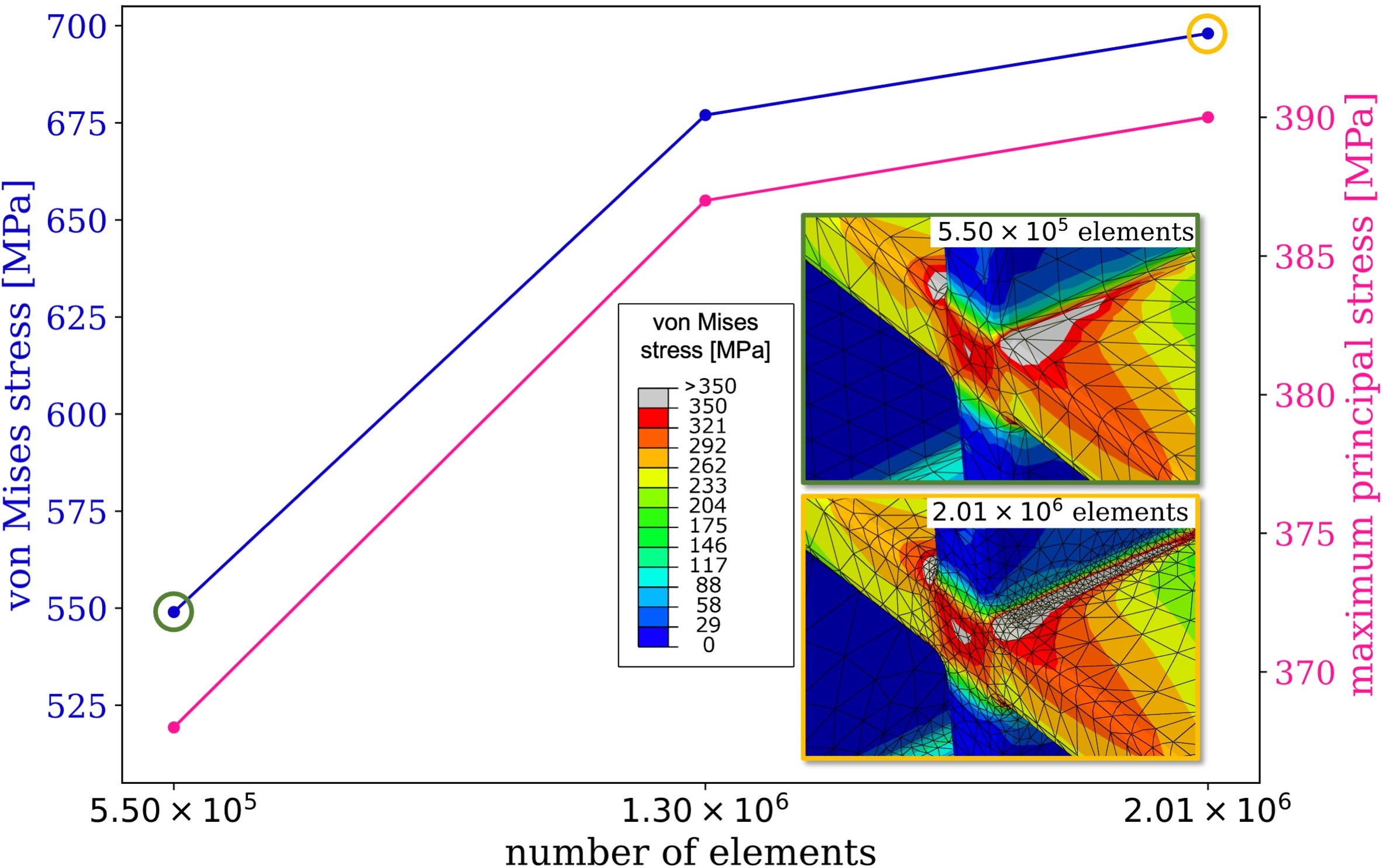}%
}\hfill
\subfloat[\label{fig:convergenge_2}]{%
  \includegraphics[width=0.48\textwidth]{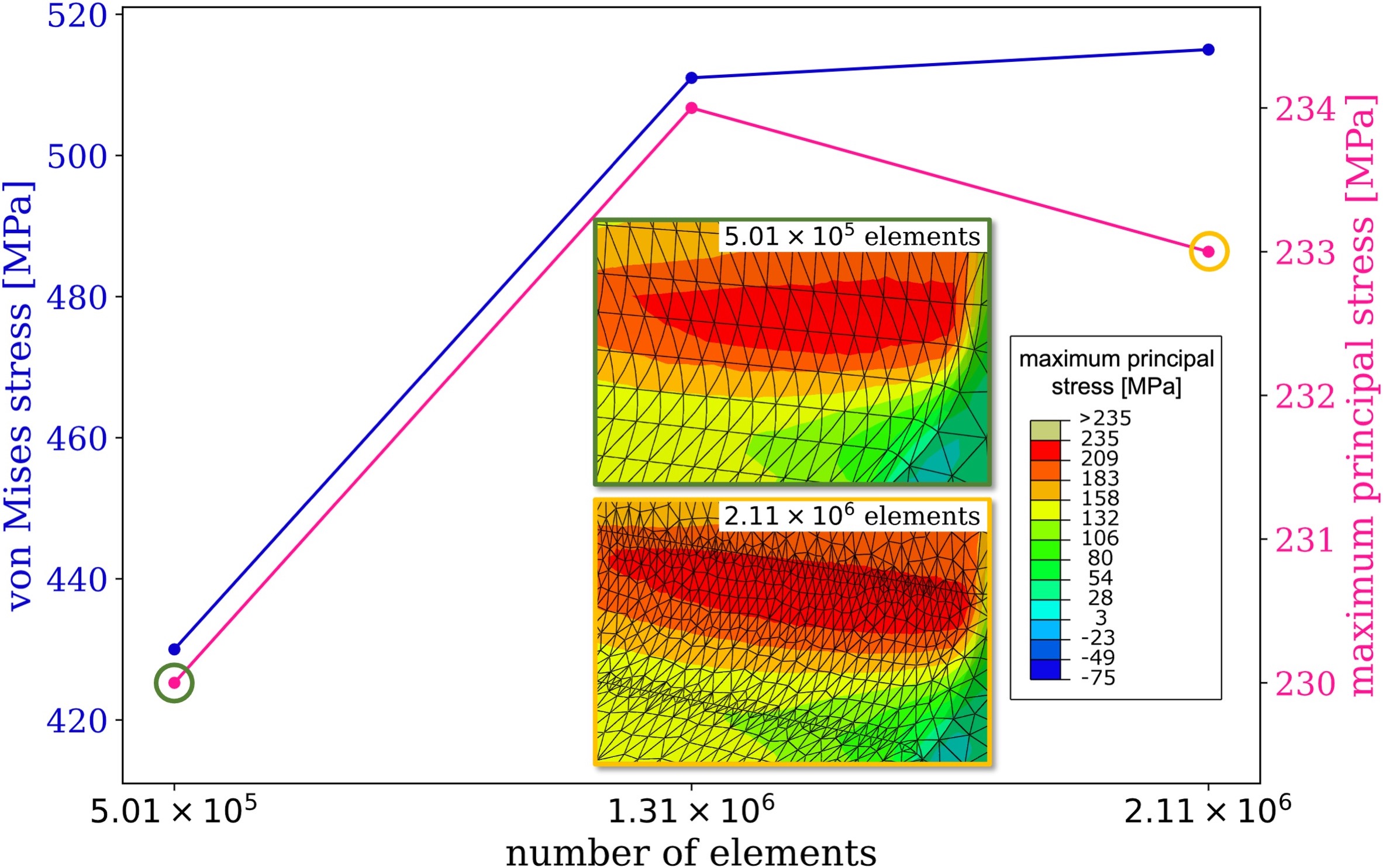}%
}\hfill
    \caption{Variation of the von Mises and maximum principal stresses as functions of the total number of elements in an FEM model for (a) the initial design and (b) the modified design investigated in Sec.~\ref{sec:BB}. Stresses are evaluated in the regions showed at higher magnification in figures \ref{fig:BB_stresses} and \ref{fig:BB_design2}. FEM mesh in these regions is shown here with a map of (a) the von Mises and (b) the maximum principal stress, to show the effect of two iterations of the mesh refinement algorithm.}
    \label{fig:convergence}
\end{figure}
As opposed to the analysis given in the sections preceding Sec. \ref{sec:BB}, there is no analytical solution for the breeding blanket module described in Sec.~\ref{sec:BB} to assess the quality of the FEM mesh. To obtain the numerical FEM results presented in  Sec.~\ref{sec:BB}, we started from a relatively coarse mesh of about $5\times10^5$ tetrahedral elements and applied to the entire model the adaptive remeshing algorithm of Abaqus, reaching a final mesh of about $2\times10^6$ elements. In such a way, the program was allowed to decrease the element size down to about \SI{0.1}{mm} at locations where the gradients of stress were relatively high, at the same time coarsening the mesh where gradients were relatively small.

Fig.~\ref{fig:convergence_1} shows the highest values of the von Mises and maximum principal stresses as a function of the total number of elements included in the FEM representation of the first design with horizontal and vertical inner walls. The mesh refinement procedure was stopped after two iterations as the change in these highest stresses was considered to be acceptably small (3\% for the von Mises, 1\% for the maximum principal stress). A similar mesh refinement process was carried out for the modified design model with slanted inner walls and increased fillet radii, as shown in Fig.~\ref{fig:convergenge_2}. Since the region where the maximum principal stress was evaluated (i.e. the inset of Fig.~\ref{fig:BB_design2_MP}) was no longer the maximum for the entire structure, the second iteration caused the value to decrease slightly. The numerical values of stress converged to within 1\% both for the von Mises and for the maximum principal stress.

\end{appendix}

\bibliography{reference}


\end{document}